\UseRawInputEncoding
\documentclass[aip,jcp,reprint,twocolumn,superscriptaddress,showpacs,floatfix]{revtex4-1}

\usepackage{epsfig}
\usepackage{mathrsfs}
\usepackage{graphicx}
\usepackage{amssymb}
\usepackage{multirow}
\usepackage{color}
\usepackage{dcolumn}
\usepackage{bm}
\usepackage{float}
\usepackage{placeins}
\definecolor{red}{rgb}{1,0,0}

\newcommand{\beq}{\begin{equation}}
\newcommand{\eeq}{\end{equation}}

\begin{document}

\newcommand{\ccpq}[2]{CC(\textit{#1};\textit{#2})}

\preprint{J. E. Deustua et al., submitted to J. Chem. Phys.}

\title{High-Level Coupled-Cluster
Energetics by Monte Carlo Sampling and Moment Expansions:
Further Details and Comparisons}

\author{J. Emiliano Deustua}
\affiliation{Department of Chemistry,
Michigan State University, East Lansing, Michigan 48824, USA}

\author{Jun Shen}
\affiliation{Department of Chemistry,
Michigan State University, East Lansing, Michigan 48824, USA}

\author{Piotr Piecuch}
\thanks{Corresponding author}
\email[e-mail: ]{piecuch@chemistry.msu.edu.}
\affiliation{Department of Chemistry,
Michigan State University, East Lansing, Michigan 48824, USA}
\affiliation{Department of Physics and Astronomy,
Michigan State University, East Lansing, Michigan 48824, USA}

\date{\today}

\begin{abstract}
We recently proposed a novel approach to converging electronic energies equivalent to
high-level coupled-cluster (CC) computations by combining the deterministic
CC($P$;$Q$) formalism with the stochastic configuration interaction (CI) and CC
Quantum Monte Carlo (QMC) propagations. This article extends our initial study
[J. E. Deustua, J. Shen, and P. Piecuch, {\it Phys. Rev. Lett.} {\bf 119}, 223003 (2017)],
which focused on recovering the energies obtained with the CC method with singles, doubles,
and triples (CCSDT) using the information extracted from full CI QMC and CCSDT-MC, to the
CIQMC approaches truncated at triples and quadruples. It also reports our first
semi-stochastic CC($P$;$Q$) calculations aimed at converging the energies that correspond
to the CC method with singles, doubles, triples, and quadruples (CCSDTQ). The ability of
the semi-stochastic CC($P$;$Q$) formalism to recover the CCSDT and CCSDTQ energies, even when
electronic quasi-degeneracies and triply and quadruply excited clusters become substantial,
is illustrated by a few numerical examples, including the F--F bond breaking in ${\rm F}_{2}$, the automerization of
cyclobutadiene, and the double dissociation of the water molecule.
\end{abstract}

\maketitle

\section{Introduction}
\label{sec1}

One of the main goals of quantum chemistry is to provide an accurate and
systematically improvable description of many-electron correlation effects
needed to determine molecular potential energy and property surfaces and
understand chemical reactivity and various types of spectroscopy.
In searching for the best solutions in this area, the size extensive methods based
on the exponential wave function ansatz \cite{Hubbard:1957,Hugenholtz:1957}
of coupled-cluster (CC) theory,\cite{Coester:1958,Coester:1960,cizek1,cizek2,cizek4}
\beq
|\Psi \rangle = e^{T} |\Phi \rangle ,
\label{eq-ccansatz}
\eeq
where
\beq
T = \sum_{n=1}^{N} T_{n}
\label{eq-clusterop}
\eeq
is the cluster operator, $T_{n}$ is the $n$-body component of $T$, $N$ is the number of
correlated electrons, and $|\Phi\rangle$ is the reference determinant, and
their extensions to excited, open-shell, and multi-reference states
\cite{paldus-li,succ5,bartlett-musial2007,chemrev-2012b,evangelista-perspective-jcp-2018}
are among the top contenders. In this study, we focus on the higher-rank members of the
single-reference CC hierarchy beyond the basic CC singles and doubles (CCSD) level,
where $T$ is truncated at $T_{2}$,\cite{ccsd,ccsd2,ccsdfritz,osaccsd} especially on the CC
approach with singles, doubles, and triples (CCSDT), where $T$ is truncated
at $T_{3}$,\cite{ccfullt,ccfullt2,ch2-bartlett2} and the CC approach with singles,
doubles, triples, and quadruples (CCSDTQ), where $T$ is truncated at $T_{4}$.
\cite{ccsdtq0,ccsdtq1,ccsdtq2} This is motivated by the fact that in great many cases
relevant to chemistry, including molecular properties at equilibrium geometries,
multi-reference situations involving smaller numbers of strongly correlated electrons, as in the
case of bond breaking and formation in the course of chemical reactions, noncovalent interactions, and
photochemistry, the single-reference CCSD, CCSDT, CCSDTQ, etc. methods and their equation-of-motion (EOM)
\cite{emrich,eomcc1,eomcc3,eomccsdt1,eomccsdt2,eomccsdt3,kallaygauss,hirata1} and linear response
\cite{monk,monk2,mukherjee_lrcc,sekino-rjb-1984,lrcc3,lrcc4,jorgensen,kondo-1995,kondo-1996}
extensions rapidly converge to the exact, full configuration interaction (FCI) limit, allowing one
to incorporate the relevant many-electron correlation effects in a conceptually straightforward
manner through particle-hole excitations from a single Slater determinant defining the Fermi vacuum
without loss of accuracy as the system becomes larger characterizing truncated CI methods.
\cite{bartlett-musial2007}

The convergence of the single-reference CCSD, CCSDT, CCSDTQ, etc. hierarchy toward FCI
in situations other than larger numbers of strongly entangled electrons is fast, but costs
of the post-CCSD computations needed to achieve a quantitative description, which are
determined by the iterative $n_{o}^{3} n_{u}^{5}$ steps in the CCSDT case and the iterative
$n_{o}^{4} n_{u}^{6}$ steps in the case of CCSDTQ, where $n_{o}$ ($n_{u}$) is the number
of occupied (unoccupied) correlated orbitals, are usually prohibitively expensive. This
is why part of the CC method development effort has been devoted to finding
approximate ways of incorporating higher--than--two-body components of the cluster operator
$T$, i.e., $T_{n}$ components with $n > 2$, and the analogous higher-order components of
the EOMCC excitation, electron-attachment, and electron-detachment operators, which could
reduce enormous computational costs of the CCSDT, CCSDTQ, and similar schemes, while eliminating
failures of the CCSD[T],\cite{ref:24a} CCSD(T),\cite{ccsdpt} CCSDT-1,\cite{ref:21a,ref:21b}
CC3,\cite{cc3_1,cc3_2} and other perturbative CC approaches (cf. Ref. \onlinecite{bartlett-musial2007}
for a review) that fail when bond breaking, biradicals, and other typical multi-reference situations
in chemistry are examined.\cite{paldus-li,bartlett-musial2007,irpc,PP:TCA,piecuch-qtp} In fact, the
analogous effort has been taking place in other areas of many-body theory, such as studies of nuclear
matter, where a systematic, computationally efficient, and robust incorporation of higher-order
many-particle correlation effects is every bit as important as in the case of electronic structure theory
and where the quantum-chemistry-inspired CC and EOMCC methods, thanks, in part, to our group's involvement,
\cite{nuclei1,nuclei5,nuclei6,nuclei8,nuclei9,nuclei10,nuclei12} have become quite popular (see, e.g.,
Ref. \onlinecite{hagen2014} and references therein). While substantial progress in the above area,
reviewed, for example, in Refs. \onlinecite{bartlett-musial2007,piecuch-qtp,jspp-chemphys2012},
has already been made, the search for the optimum solution that would allow us to obtain the
results of the full CCSDT, full CCSDTQ, or similar quality at the fraction of the cost and
without having to rely on perturbative concepts or user- and system-dependent ideas, such as
the idea of active orbitals to select higher--than--two-body components of the cluster and EOMCC
excitation operators,\cite{piecuch-qtp} continues.

In order to address this situation, we have started exploring a radically
new way of converging accurate electronic energetics equivalent to those obtained
with the high-level CC approaches of the full CCSDT, full CCSDTQ, and similar types,
at the small fraction of the computational cost and preserving the black-box character
of conventional single-reference methods, even when higher--than--two-body components
of the cluster and excitation operators characterizing potential energy surfaces
along bond stretching coordinates become large.\cite{stochastic-ccpq-prl-2017} The
key idea of the approach suggested in Ref. \onlinecite{stochastic-ccpq-prl-2017}, which
we have recently extended to excited states,\cite{eomccp-jcp-2019,stochastic-ccpq-molphys-2020}
is a merger of the deterministic formalism, abbreviated as CC($P$;$Q$),
\cite{jspp-chemphys2012,jspp-jcp2012,jspp-jctc2012,nbjspp-molphys2017}
which enables one to correct energies obtained with conventional as well as unconventional
truncations in the cluster and EOMCC excitation operators for any category of many-electron
correlation effects of interest, with the stochastic FCI Quantum Monte Carlo (FCIQMC)
\cite{Booth2009,Cleland2010,fciqmc-uga-2019,ghanem_alavi_fciqmc_jcp_2019}
and CC Monte Carlo (CCMC)\cite{Thom2010,Franklin2016,Spencer2016,Scott2017} methods
(cf. Refs. \onlinecite{cad-fciqmc-jcp-2018,vitale-alavi-kats-jctc-2020,eriksen-et-al-jpcl-2020} for alternative ways
of combining FCIQMC with the deterministic CC framework).
As shown in Refs. \onlinecite{stochastic-ccpq-prl-2017,stochastic-ccpq-molphys-2020}, where we reported
preliminary calculations aimed at recovering full CCSDT and EOMCCSDT\cite{,eomccsdt1,eomccsdt2,eomccsdt3}
energetics, the resulting
semi-stochastic CC($P$;$Q$) methodology, using the FCIQMC and CCSDT-MC approaches to identify
the leading determinants or cluster amplitudes in the wave function and the {\it a posteriori}
CC($P$;$Q$) corrections to capture the remaining correlations, rapidly converges to the
target energetics based on the information extracted from the early stages of FCIQMC or CCSDT-MC
propagations. If confirmed through additional tests and comparisons involving various QMC and CC levels,
the merger of the deterministic CC($P$;$Q$) and stochastic CIQMC and CCMC ideas, originally proposed in
Ref. \onlinecite{stochastic-ccpq-prl-2017}, may substantially impact accurate quantum calculations for
many-electron and other many-fermion systems, opening interesting new possibilities in this area.

The present study is our next step in the development and examination of the semi-stochastic CC($P$;$Q$)
methodology. In this work, we extend our initial study,\cite{stochastic-ccpq-prl-2017}
which focused on recovering the full CCSDT energetics based on the information extracted
from the FCIQMC and CCSDT-MC propagations, to the CIQMC methods truncated at triples
(CISDT-MC) or triples and quadruples (CISDTQ-MC), which may offer significant savings
in the computational effort compared to FCIQMC and which are formally compatible with
the CCSDT and CCSDTQ excitation manifolds we would like to capture. We also report
our initial results of the semi-stochastic
CC($P$;$Q$) calculations aimed at converging the full CCSDTQ energetics.
The ability of the semi-stochastic CC($P$;$Q$) approaches to recover the CCSDT and
CCSDTQ energies based on the truncated CISDT-MC and CISDTQ-MC propagations,
even when electronic quasi-degeneracies and $T_{3}$ and $T_{4}$
clusters become substantial, is illustrated using the challenging cases of the
F--F bond breaking in ${\rm F}_{2}$, the automerization of cyclobutadiene, and the double dissociation
of the water molecule as examples.

\section{Theory and Algorithmic Details}
\label{sec2}

As pointed out in the Introduction, the semi-stochastic CC($P$;$Q$) approach proposed
in Ref. \onlinecite{stochastic-ccpq-prl-2017} is based on combining the
deterministic CC($P$;$Q$) framework, developed mainly in Refs.
\onlinecite{jspp-chemphys2012,jspp-jcp2012,nbjspp-molphys2017},
with the CIQMC and CCMC ideas that were originally laid down in Refs.
\onlinecite{Booth2009,Cleland2010,Thom2010}. Thus, we divide this section into two subsections.
In Section \ref{sec2.1}, we summarize the key elements of the deterministic CC($P$;$Q$) formalism,
focusing on the ground-state problem relevant to the calculations reported in this
study. Section \ref{sec2.2} provides information about the semi-stochastic
CC($P$;$Q$) methods developed and tested in this work, which aim at converging
the CCSDT and CCSDTQ energies with the help of the FCIQMC, CISDT-MC, and CISDTQ-MC
propagations.

\subsection{Basic Elements of the Ground-State
CC(\protect\mbox{\boldmath$P;Q$}) Formalism}
\label{sec2.1}

The CC($P$;$Q$) formalism has emerged out of our interest in generalizing
the biorthogonal moment energy expansions, which in the past resulted in the
completely renormalized (CR) CC and EOMCC approaches, including CR-CC(2,3),
\cite{crccl_jcp,crccl_cpl,crccl_molphys,crccl_jpc,crccl_ijqc} CR-EOMCC(2,3),
\cite{crccl_molphys,crccl_ijqc2} $\delta$-CR-EOMCC(2,3),\cite{7hq} and their higher-order extensions,
\cite{nbjspp-molphys2017,ccpq-be2-jpca-2018,ptcp2007,msg65,nuclei8} such that one
can correct the CC/EOMCC energies obtained with unconventional truncations in the
cluster and EOMCC excitation operators, in addition to the
conventional ones at a given many-body rank,
for essentially any category of many-electron correlation effects of interest.
The CC($P$;$Q$) framework is general, i.e., it applies to ground as well as excited states,
but since this work deals with the calculations that aim at recovering
the ground-state CCSDT and CCSDTQ energetics, in the description below
we focus on the ground-state CC($P$;$Q$) theory.

According to the formal CC($P$;$Q$) prescription, the ground-state energy of a $N$-electron
system is determined in two steps. In the initial, iterative, CC($P$) step, we solve
the CC equations in the subspace ${\mathscr H}^{(P)}$ of the $N$-electron Hilbert space
${\mathscr H}$. We assume that subspace ${\mathscr H}^{(P)}$, which we also call the $P$ space,
is spanned by the excited determinants $|\Phi_{K}\rangle = E_{K} |\Phi\rangle$
that together with the reference determinant $|\Phi\rangle$ provide the leading contributions
to the target ground state $|\Psi\rangle$ ($E_{K}$ designates the usual
elementary particle-hole excitation operator generating $|\Phi_{K}\rangle$ from $|\Phi\rangle$).
In other words, we approximate the cluster operator $T$ in Eq. (\ref{eq-ccansatz}) by
\beq
T^{(P)} = \sum_{|\Phi_{K} \rangle \in {\mathscr H}^{(P)}} t_{K} E_{K}
\label{eq:tp}
\eeq
and solve the usual system of CC equations,
\beq
{\mathfrak M}_{K}(P) = 0, \;\; |\Phi_{K} \rangle \in {\mathscr H}^{(P)},
\label{eq:cceqs}
\eeq
where
\beq
{\mathfrak M}_{K}(P) = \langle \Phi_{K} |\bar{H}^{(P)}|\Phi \rangle
\label{eq:mom}
\eeq
are the generalized moments of the $P$-space CC equations\cite{moments,leszcz,ren1} and
\beq
\bar{H}^{(P)} = e^{-T^{(P)}}He^{T^{(P)}} = (He^{T^{(P)}})_{C}
\label{eq:hbar}
\eeq
is the relevant similarity-transformed Hamiltonian, for the cluster amplitudes $t_{K}$
(subscript $C$ in Eq. (\ref{eq:hbar}) designates the connected operator product).
Once the cluster operator $T^{(P)}$ and the ground-state energy
\beq
E^{(P)} = \langle \Phi | \bar{H}^{(P)} |\Phi\rangle
\label{eq:ccpenergy}
\eeq
that corresponds to it are determined,
we proceed to the second step of CC($P$;$Q$) considerations, which is the calculation of the
noniterative correction $\delta(P;Q)$ to the CC($P$) energy $E^{(P)}$ that accounts for the
many-electron correlation effects captured by another subspace of the $N$-electron Hilbert
space ${\mathscr H}$, designated as ${\mathscr H}^{(Q)}$ and called the $Q$ space, which satisfies
the condition ${\mathscr H}^{(Q)} \subseteq ({\mathscr H}^{(0)} \oplus {\mathscr H}^{(P)})^{\perp}$,
where ${\mathscr H}^{(0)}$ is a one-dimensional subspace of ${\mathscr H}$ spanned by the
reference determinant $|\Phi\rangle$. The formula for the $\delta(P;Q)$ correction is
\cite{jspp-chemphys2012,stochastic-ccpq-prl-2017,stochastic-ccpq-molphys-2020,jspp-jcp2012,nbjspp-molphys2017}
\beq
\delta(P;Q) =
\sum_{
\begin{array}{c}{\scriptstyle
|\Phi_{K}\rangle \in {\mathscr H}^{(Q)}}
\\ [-1mm]
{\scriptstyle {\rm rank}(|\Phi_{K}\rangle) \leq \min(N_{0}^{(P)},\Xi^{(Q)})}
\end{array}
}
\ell_{K}(P) \; {\mathfrak M}_{K}(P) ,
\label{mmcc-gen-delta}
\eeq
where integer $N^{(P)}$ defines the highest many-body rank of the excited determinants
$|\Phi_{K}\rangle$ relative to $|\Phi\rangle$ (${\rm rank}(|\Phi_{K}\rangle)$) for
which moments ${\mathfrak M}_{K}(P)$, Eq. (\ref{eq:mom}), are still non-zero and $\Xi^{(Q)}$ is the
highest many-body rank of the excited determinant(s) $|\Phi_{K}\rangle$
included in ${\mathscr H}^{(Q)}$. In practical CC($P$;$Q$) calculations, including those
discussed in Section \ref{sec3}, the $\ell_{K}(P)$ coefficients entering Eq. (\ref{mmcc-gen-delta})
are calculated as
\beq
\ell_{K}(P) = \langle \Phi | ({\bf 1} + {\Lambda}^{(P)}) \bar{H}^{(P)} |\Phi_{K}\rangle/D_{K}(P),
\label{eq:ell}
\eeq
where ${\bf 1}$ is the unit operator,
\beq
{\Lambda}^{(P)} = \sum_{|\Phi_{K} \rangle \in {\mathscr H}^{(P)}} {\lambda}_{K} (E_{K})^{\dagger}
\label{eq:lambda}
\eeq
is the hole-particle deexcitation operator defining the bra state
$\langle\tilde{\Psi}^{(P)} | = \langle \Phi | (1 + {\Lambda}^{(P)}) e^{-T^{(P)}}$
corresponding to the CC($P$) ket state $|\Psi^{(P)}\rangle = e^{T^{(P)}} |\Phi\rangle$, and
\beq
D_{K}(P) = E^{(P)} - \langle \Phi_{K} | \bar{H}^{(P)} | \Phi_{K}\rangle.
\label{eq:denom}
\eeq
One determines ${\Lambda}^{(P)}$, or the amplitudes ${\lambda}_{K}$ that define it, by
solving the linear system of equations representing the left eigenstate CC problem
\cite{bartlett-musial2007} in the $P$ space, i.e.,
\beq
\langle \Phi | ({\bf 1} + {\Lambda}^{(P)}) \bar{H}^{(P)} | \Phi_{K} \rangle = E^{(P)} {\lambda}_{K},
\;\;
|\Phi_{K} \rangle \in {\mathscr H}^{(P)},
\label{left-cc-gen}
\eeq
where $E^{(P)}$ is the previously determined CC($P$) energy. Once the noniterative
correction $\delta(P;Q)$ is determined, the CC($P$;$Q$) energy is obtained as
\beq
E^{(P+Q)} = E^{(P)} + \delta(P;Q) .
\label{mmcc-gen}
\eeq
In practice, we often distinguish between the complete version of the CC($P$;$Q$) theory, designated,
following Refs. \onlinecite{stochastic-ccpq-prl-2017,jspp-jctc2012}, as CC($P$;$Q$)$_{\rm EN}$, which
uses the Epstein--Nesbet-like denominator $D_{K}(P)$, Eq. (\ref{eq:denom}), in calculating the
$\ell_{K}(P)$ amplitudes, and the approximate version of CC($P$;$Q$), abbreviated as CC($P$;$Q$)$_{\rm MP}$,
which relies on the M{\o}ller--Plesset form of $D_{K}(P)$ obtained by replacing $\bar{H}^{(P)}$
in Eq. (\ref{eq:denom}) by the bare Fock operator (cf., e.g., Refs.
\onlinecite{stochastic-ccpq-prl-2017,jspp-jctc2012,nbjspp-molphys2017}). Both of these
variants of the CC($P$;$Q$) formalism are considered in this study.

We must now come up with the appropriate choices of the $P$ and $Q$ spaces entering the CC($P$;$Q$)
considerations that would allow us to match the quality of the high-level CC computations
of the CCSDT, CCSDTQ, and similar type at the small fraction of the cost. As is often the case
in the CC work, one could start from the conventional choices,
where the $P$ space ${\mathscr H}^{(P)}$ is spanned by all excited
$|\Phi_{i_{1} \ldots i_{n}}^{a_{1} \ldots a_{n}} \rangle$ determinants with the excitation rank
$n \leq m_{A}$, where $i_{1}, i_{2}, \ldots$ ($a_{1}, a_{2}, \ldots$)
designate the spin-orbitals occupied (unoccupied) in $|\Phi\rangle$, and the $Q$ space
${\mathscr H}^{(Q)}$ by those with $m_{A} < n \leq m_{B}$, where $m_{B} \leq N$. In that case,
one ends up with the well-stablished CR-CC($m_{A}$,$m_{B}$) hierarchy,
\cite{jspp-chemphys2012,crccl_jcp,crccl_cpl,crccl_molphys,crccl_jpc,crccl_ijqc,crccl_ijqc2,%
nbjspp-molphys2017,ptcp2007,nuclei8}
including the aforementioned CR-CC(2,3) approximation, where $m_{A} = 2$ and $m_{B} = 3$,
and the related CCSD(2)$_{\rm T}$ \cite{ccsdpt2}
(cf., also, Refs. \onlinecite{gwaltney1,eomccpt,gwaltney3}), CCSD(T)$_{\Lambda}$,\cite{stanton1997,crawford1998,ref:26}
$\Lambda$-CCSD(T),\cite{bartlett2008a,bartlett2008b}
and similar\cite{irpc,PP:TCA,leszcz,ren1,ndcmmcc} schemes that allow one to correct the CCSD energies for triples.
The CR-CC(2,3) method is useful, improving, for example, poor performance of CCSD(T) in covalent bond breaking
situations \cite{jspp-chemphys2012,crccl_jcp,crccl_cpl,crccl_molphys,ptcp2007,crccl_jpc,ge1,ge2} and for certain
classes of noncovalent interactions\cite{ccpq-be2-jpca-2018,ccpq-mg2-mp-2019} without a substantial increase
of the computational effort, but neither CR-CC(2,3) nor its CCSD(2)$_{\rm T}$, CCSD(T)$_{\Lambda}$, and
$\Lambda$-CCSD(T) counterparts (which are all approximations to CR-CC(2,3)) are free from drawbacks.
One of the main problems with CR-CC(2,3), CCSD(2)$_{\rm T}$,
$\Lambda$-CCSD(T), and other noniterative corrections to CCSD is the fact that, in analogy to CCSD(T), they
decouple the higher-order $T_{n}$ components with $n > m_{A}$, such as $T_{3}$ or $T_{3}$ and $T_{4}$, from
their lower-order $n \leq m_{A}$ (e.g., $T_{1}$ and $T_{2}$) counterparts. This can result in substantial
errors, for example when the activation energies and chemical reaction profiles involving rearrangements
of $\pi$ bonds and singlet--triplet gaps in certain classes of biradical species are examined.
\cite{jspp-jcp2012,jspp-jctc2012,nbjspp-molphys2017} The automerization of cyclobutadiene, which is one of
the benchmark examples in Section \ref{sec3}, provides an illustration of the challenges the noniterative
corrections to CCSD, including CCSD(2)$_{\rm T}$ and CR-CC(2,3), face when the coupling of the lower-order
$T_{1}$ and $T_{2}$ and higher-order $T_{3}$ clusters becomes significant (see Ref. \onlinecite{jspp-jcp2012}
for further analysis and additional remarks). One can address problems of this type by using active orbitals
to incorporate the dominant higher--than--doubly excited determinants, in addition to all singles and doubles,
in the $P$ space, as in the successful CC(t;3), CC(t,q;3), and CC(t,q;3,4) hierarchy,
\cite{jspp-chemphys2012,jspp-jcp2012,jspp-jctc2012,nbjspp-molphys2017,ccpq-be2-jpca-2018,ccpq-mg2-mp-2019}
which uses the CC($P$;$Q$) framework to correct the results of the active-space CCSDt
\cite{piecuch-qtp,semi0b,semi2,ccsdtq3,semih2o,ghose,semi3c,semi4,semi4new} or CCSDtq
\cite{piecuch-qtp,semi0a,semi2,ccsdtq3,semi4} calculations for the remaining $T_{3}$ or $T_{3}$ and $T_{4}$
correlations that were not captured via active orbitals, but the resulting methods are no longer
computational black boxes. The semi-stochastic CC($P$;$Q$) methodology, introduced in Ref.
\onlinecite{stochastic-ccpq-prl-2017}, extended to excited states in Ref. \onlinecite{stochastic-ccpq-molphys-2020},
and further developed in this work, which takes advantage of the FCIQMC or truncated CIQMC/CCMC propagations
that can identify the leading higher--than--doubly-excited determinants for the inclusion in the $P$ space,
while using the noniterative $\delta(P;Q)$ corrections to capture the remaining correlations of interest,
offers an automated way of performing CC($P$;$Q$) computations without any reference to the
user- and system-dependent active orbitals. The semi-stochastic CC($P$;$Q$) methods developed and
tested in this study are discussed next.

\subsection{Semi-stochastic CC(\protect\mbox{\boldmath$P;Q$}) Approaches Using FCIQMC and
its Truncated CISDT-MC and CISDTQ-MC Counterparts}
\label{sec2.2}

In our original examination of the semi-stochastic CC($P$;$Q$) framework\cite{stochastic-ccpq-prl-2017} and
its recent extension to excited states,\cite{stochastic-ccpq-molphys-2020} where we focused on converging
the full CCSDT and EOMCCSDT energetics, we demonstrated that the FCIQMC and CCSDT-MC approaches are capable
of generating meaningful $P$ spaces for the subsequent CC($P$)/EOMCC($P$) iterations, which precede the
determination of the $\delta(P;Q)$ moment corrections, already in the early stages of the respective QMC
propagations. The main objective of this work is to explore if the same remains true when FCIQMC is
replaced by its less expensive truncated CISDT-MC and CISDTQ-MC counterparts, in which spawning beyond
the triply excited (CISDT-MC) or quadruply excited (CISDTQ-MC) determinants is disallowed, and if one
can use the CIQMC-driven CC($P$;$Q$) calculations to converge the higher-level CCSDTQ energetics with
similar efficiency.

The key steps of the semi-stochastic CC($P$;$Q$) algorithm exploited in this study, which allows us to
converge the CCSDT and CCSDTQ energetics using the $P$ spaces extracted from the FCIQMC and truncated
CISDT-MC and CISDTQ-MC propagations, are as follows:
\begin{itemize}
\item[1.]
Initiate a CIQMC run appropriate for the CC method of interest by placing a certain number of walkers
on the reference state $|\Phi\rangle$, which in all of the calculations reported in this article is
the restricted Hartree-Fock (RHF) determinant. Among the CIQMC schemes that can provide meaningful $P$
spaces for the CC($P$;$Q$) calculations targeting the CCSDT energetics are the FCIQMC approach used
in our earlier work\cite{stochastic-ccpq-prl-2017,eomccp-jcp-2019,stochastic-ccpq-molphys-2020} and
the CISDT-MC and CISDTQ-MC methods examined in the present study.
If the objective is to converge the CCSDTQ energetics, one can use FCIQMC or CISDTQ-MC, which are the
two choices pursued in the present work, but not CISDT-MC, which ignores quadruply excited determinants.
As in our earlier semi-stochastic CC($P$)/EOMCC($P$) and
CC($P$;$Q$) work,\cite{stochastic-ccpq-prl-2017,eomccp-jcp-2019,stochastic-ccpq-molphys-2020}
all of the calculations reported in this article adopted the initiator CIQMC ($i$-CIQMC) algorithm,
originally proposed in Ref. \onlinecite{Cleland2010}, based on integer walker numbers,
but the procedure discussed here is flexible and could be merged with other
CIQMC techniques developed in recent years, such as those described in Refs.
\onlinecite{ghanem_alavi_fciqmc_jcp_2019,ghanem_alavi_fciqmc_2020}.
\item[2.]
After a certain number of CIQMC time steps, called MC iterations, i.e., after some QMC propagation time $\tau$,
extract a list of higher--than--doubly excited determinants relevant to the CC theory of interest to construct
the $P$ space for executing the CC($P$) calculations. If one is interested in targeting the CCSDT-level energetics,
the $P$ space used in the CC($P$) iterations consists of
all singly and doubly excited determinants and a subset of triply excited determinants identified by the
underlying FCIQMC, CISDT-MC, or CISDTQ-MC propagation, where each triply excited determinant in the subset
is populated by at least $n_{P}$ positive or negative walkers. In analogy to our previous CC($P$)/EOMCC($P$)
and CC($P$;$Q$) studies,\cite{stochastic-ccpq-prl-2017,eomccp-jcp-2019,stochastic-ccpq-molphys-2020}
all of the CC($P$) and CC($P$;$Q$) computations carried out in this work use $n_{P} = 1$. If the goal is to
converge the CCSDTQ energetics, the $P$ space for the CC($P$) computations is defined as all singly and doubly
excited determinants and a subset of triply and quadruply excited determinants identified by the underlying FCIQMC
or CISDTQ-MC propagation, where, again, each triply and quadruply excited determinant in the subset is populated
by a minimum of $n_{P}$ positive or negative walkers.
\item[3.]
Solve the CC($P$) and left-eigenstate CC($P$) equations, Eqs. (\ref{eq:cceqs}) and (\ref{left-cc-gen}),
respectively, where $E^{(P)}$ is given by Eq. (\ref{eq:ccpenergy}), for the cluster operator $T^{(P)}$ and
the deexcitation operator ${\Lambda}^{(P)}$ in the $P$ space determined in step 2. If the objective
is to converge the CCSDT-level energetics, we define $T^{(P)} = T_{1} + T_{2} + T_{3}^{\rm (MC)}$ and
${\Lambda}^{(P)} = \Lambda_{1} + \Lambda_{2} + \Lambda_{3}^{\rm (MC)}$, where $T_{3}^{\rm (MC)}$ and
$\Lambda_{3}^{\rm (MC)}$ are the three-body components of $T^{(P)}$ and ${\Lambda}^{(P)}$, respectively,
defined using the list of triples identified by the FCIQMC, CISDT-MC, or CISDTQ-MC propagation at time $\tau$,
as described in step 2.
If one is targeting the CCSDTQ-level energetics, $T^{(P)} = T_{1} + T_{2} + T_{3}^{\rm (MC)} + T_{4}^{\rm (MC)}$
and ${\Lambda}^{(P)} = \Lambda_{1} + \Lambda_{2} + \Lambda_{3}^{\rm (MC)} + \Lambda_{4}^{\rm (MC)}$,
where $T_{3}^{\rm (MC)}$ and $\Lambda_{3}^{\rm (MC)}$ are the three-body and $T_{4}^{\rm (MC)}$ and
$\Lambda_{4}^{\rm (MC)}$ four-body components of $T^{(P)}$ and ${\Lambda}^{(P)}$, respectively, defined using
the lists of triples and quadruples identified by the FCIQMC or CISDTQ-MC propagation at time $\tau$.
\item[4.]
Use the CC($P$;$Q$) correction $\delta(P;Q)$, Eq. (\ref{mmcc-gen-delta}), to correct the energy $E^{(P)}$
obtained in step 3 for the remaining correlation effects of interest, meaning those correlations that were
not captured by the CC($P$) calculations performed at the time $\tau$ the list of higher--than--doubly
excited determinants entering the relevant $P$ space was created.
If the objective is to converge the CCSDT-level energetics, the $Q$ space entering the definition of
$\delta(P;Q)$ consists of those triply excited determinants that in the FCIQMC, CISDT-MC, or CISDTQ-MC
propagation at time $\tau$ are populated by less than $n_{P}$ positive or negative walkers (in this study,
where $n_{P} = 1$, the triply excited determinants that were not captured by the FCIQMC, CISDT-MC, or
CISDTQ-MC propagation at time $\tau$). If the goal is to recover the CCSDTQ-level energetics, the
$Q$ space used to calculate $\delta(P;Q)$ consists of the triply and quadruply excited determinants
that in the FCIQMC or CISDTQ-MC propagation at time $\tau$ are populated by less than $n_{P}$ positive
or negative walkers.
\item[5.]
Check the convergence of the CC($P$;$Q$) energy $E^{(P+Q)}$, Eq. (\ref{mmcc-gen}), obtained in step 4, by
repeating steps 2--4 at some later CIQMC propagation time $\tau^{\prime} > \tau$. If the resulting energy
$E^{(P+Q)}$ no longer changes within a given convergence threshold, the CC($P$;$Q$) calculation can be stopped.
As pointed out in Refs. \onlinecite{stochastic-ccpq-prl-2017,eomccp-jcp-2019,stochastic-ccpq-molphys-2020}, one
can also stop it once the fraction (fractions) of higher--than--doubly excited determinants captured by the
CIQMC propagation relevant to the target CC theory level, included in the $P$ space, is (are) sufficiently
large to obtain the desired accuracy. This is further discussed in Section \ref{sec3}, where the numerical
results obtained in this study are presented.
\end{itemize}

The above semi-stochastic CC($P$;$Q$) algorithm, allowing us to recover the CCSDT and CCSDTQ energetics
using the $P$ spaces identified with the help of FCIQMC or truncated CISDT-MC and CISDTQ-MC propagations,
has been implemented by modifying our previously developed standalone deterministic CC($P$;$Q$) codes,
\cite{jspp-chemphys2012,jspp-jcp2012,nbjspp-molphys2017} which rely on the RHF, restricted open-shell
Hartree-Fock, and integral routines in the GAMESS package,\cite{gamess,gamess2020} such that they could
handle the stochastically determined lists of triples and quadruples, and by interfacing the resulting
program with the $i$-CIQMC routines available in the HANDE software.\cite{hande-jors-2015,hande-jctc-2019}
As in our earlier semi-stochastic CC($P$)/EOMCC($P$) and CC($P$;$Q$) work,
\cite{stochastic-ccpq-prl-2017,eomccp-jcp-2019,stochastic-ccpq-molphys-2020} we rely on the original form
of the initiator CIQMC ($i$-CIQMC) algorithm proposed in Ref. \onlinecite{Cleland2010}, where only those
determinants that acquire walker population exceeding a preset value $n_{a}$ are allowed to attempt 
spawning new walkers onto empty determinants, but,
as already alluded to above,
one could consider interfacing our CC($P$;$Q$)
framework with the improved ways of converging CIQMC, such as the adaptive-shift method developed
in Refs. \onlinecite{ghanem_alavi_fciqmc_jcp_2019,ghanem_alavi_fciqmc_2020}.
While the choice of a specific CIQMC algorithm may not be as critical in the context of CC($P$;$Q$) considerations
as in the case of other applications of QMC techniques, since the only role CIQMC propagations in the
semi-stochastic CC($P$;$Q$) calculations is to identify the leading higher--than--doubly excited
determinants for the inclusion in the $P$ space and, as shown in Section \ref{sec3} and
our previous studies,\cite{stochastic-ccpq-prl-2017,stochastic-ccpq-molphys-2020} moment corrections
$\delta(P;Q)$ are very efficient in accounting for the many-electron
correlation effects due to the remaining determinants not captured
by CIQMC, we are planning to integrate our CC($P$;$Q$) codes with the CIQMC methods described
in Refs. \onlinecite{ghanem_alavi_fciqmc_jcp_2019,ghanem_alavi_fciqmc_2020} in the future work.
It will be interesting to examine if the excellent performance of the semi-stochastic
CC($P$;$Q$) methods observed in the calculations reported in this article can be improved
further by replacing the $i$-CIQMC algorithm by better ways of converging CIQMC.

In the case of the semi-stochastic CC($P$;$Q$) codes aimed at converging the CCSDT energetics, which we have
extended in the present study by allowing them to work with the CISDT-MC and CISDTQ-MC approaches, in addition
to the previously examined FCIQMC\cite{stochastic-ccpq-prl-2017,eomccp-jcp-2019,stochastic-ccpq-molphys-2020}
and CCSDT-MC\cite{stochastic-ccpq-prl-2017} options, we follow the algorithm summarized in steps 1--5
without any alterations. In particular, all of the quantities entering Eq. (\ref{mmcc-gen-delta})
for the noniterative correction $\delta(P;Q)$ are treated in the present study fully. This is an improvement
compared to our original semi-stochastic CC($P$;$Q$) computations utilizing FCIQMC and CCSDT-MC,
reported in Ref. \onlinecite{stochastic-ccpq-prl-2017}, where we adopted an approximation in which the
three-body component $\Lambda_{3}^{\rm (MC)}$ of the deexcitation operator ${\Lambda}^{(P)}$ used to
determine amplitudes $\ell_{K}(P)$ entering Eq. (\ref{mmcc-gen-delta}) was neglected. In analogy to this
work, the similarity-transformed Hamiltonian $\bar{H}^{(P)}$, defining moments ${\mathfrak M}_{K}(P)$
and entering the linear system defined by Eq. (\ref{left-cc-gen}), which is used to determine
${\Lambda}^{(P)}$, was treated in Ref. \onlinecite{stochastic-ccpq-prl-2017} fully, i.e., $\bar{H}^{(P)}$
employed in the CC($P$;$Q$) calculations aimed at recovering the CCSDT energetics was defined as
$(H e^{T_{1}+T_{2}+T_{3}^{\rm (MC)}})_{C}$, so that the one- and two-body components of ${\Lambda}^{(P)}$
employed in Ref. \onlinecite{stochastic-ccpq-prl-2017} were properly relaxed in the presence of the
three-body component $T_{3}^{\rm (MC)}$ of the cluster operator $T^{(P)}$ obtained in the preceding
CC($P$) calculations, but $\Lambda_{3}^{\rm (MC)}$ was neglected. Although all of
our numerical tests to date indicate that this approximation has a small effect on the results
of the semi-stochastic CC($P$;$Q$) calculations utilizing full and truncated CIQMC and no effect
on our main conclusions, we no longer use it in this work. In other words, all of the calculations
reported in the present study rely on the complete representations of $\bar{H}^{(P)}$ and ${\Lambda}^{(P)}$
when constructing moments ${\mathfrak M}_{K}(P)$ and amplitudes $\ell_{K}(P)$ entering Eq. (\ref{mmcc-gen-delta}).
This means that $\bar{H}^{(P)}$ and ${\Lambda}^{(P)}$ used to determine the CC($P$;$Q$)
correction $\delta(P;Q)$ in the calculations aimed at the CCSDT energetics are defined as
$(H e^{T_{1}+T_{2}+T_{3}^{\rm (MC)}})_{C}$ and $\Lambda_{1} + \Lambda_{2} + \Lambda_{3}^{\rm (MC)}$,
respectively.

We have, however, introduced an approximation in the semi-stochastic CC($P$;$Q$) routines that are used to converge
the CCSDTQ-level energetics. Given the pilot nature of these routines, the noniterative correction $\delta(P;Q)$
that they produce corrects the $E^{(P)}$ energy, which is obtained in this case by solving the CC($P$)
equations in the space of all singles and doubles and subsets of triples and quadruples captured by
FCIQMC or CISDTQ-MC, for the remaining triples not included in the $P$ space, but the quadruples
contributions to $\delta(P;Q)$ are ignored. This approximation is acceptable, since in the $\tau = \infty$
limit, where the $P$ space contains all triples and quadruples, i.e., the corresponding $Q$ space is
empty, the uncorrected CC($P$) and partially or fully corrected CC($P$;$Q$) calculations recover the
CCSDTQ energetics. All of our tests to date, including those discussed in Section \ref{sec3}, indicate that
the convergence of the CC($P$;$Q$) computations, in which the quadruples component of $\delta(P;Q)$
is ignored, toward CCSDTQ is rapid, even when the $T_{4}$ effects become significant, so
the above approximation does not seem to have a major effect on the convergence rate, but we will
implement the full correction $\delta(P;Q)$ due to the missing triples as well as quadruples in the
future to examine if one can accelerate convergence toward CCSDTQ even further.

As explained in Refs. \onlinecite{stochastic-ccpq-prl-2017,stochastic-ccpq-molphys-2020} (cf., also, Ref.
\onlinecite{eomccp-jcp-2019}), the semi-stochastic CC($P$;$Q$) approaches of the type summarized above offer
a number of advantages. Among them are substantial savings in the computational effort compared to the
parent high-level CC theories they target and a systematic behavior of the resulting $E^{(P+Q)}$ energies
as $\tau$ approaches $\infty$. The latter feature is a direct consequence of the fact that if we follow the
definitions of the $P$ and $Q$ spaces introduced in steps 2 and 4 above, the initial, $\tau = 0$, CC($P$;$Q$)
energies are identical to those obtained with CR-CC(2,3) or CR-CC(2,4), which are approximations to
CCSDT and CCSDTQ, respectively, that account for some $T_{3}$ (CR-CC(2,3)) or $T_{3}$ and $T_{4}$
(CR-CC(2,4)) correlations. In the $\tau = \infty$ limit, the CC($P$;$Q$) energies $E^{(P+Q)}$ become
equivalent to their respective high-level CC parents, which account for the $T_{n}$ components with $n > 2$,
such as $T_{3}$ or $T_{3}$ and $T_{4}$, fully, so that the QMC propagation time $\tau$ becomes a parameter
connecting CR-CC(2,3) with CCSDT and CR-CC(2,4) with CCSDTQ. In the case of our current implementation
of the semi-stochastic CC($P$;$Q$) approach aimed at converging the CCSDTQ energetics, where the
quadruples contributions to correction $\delta(P;Q)$ are ignored, the initial, $\tau = 0$, CC($P$;$Q$)
energy is equivalent to that obtained with the CR-CC(2,3) approach, i.e., the QMC propagation time $\tau$
connects CR-CC(2,3) with CCSDTQ. When $\tau$ approaches $\infty$, the uncorrected
CC($P$) energies $E^{(P)}$ converge to their CCSDT and CCSDTQ parents as well, but the convergence toward
CCSDT and CCSDTQ is in this case slower, since the CC($P$) energies at $\tau = 0$ are equivalent to those
of CCSD, which has no information about the $T_{n}$ components with $n > 2$, and, as shown in our earlier
work,\cite{stochastic-ccpq-prl-2017,stochastic-ccpq-molphys-2020} and as
clearly demonstrated in the present study, the CC($P$;$Q$) corrections $\delta(P;Q)$ greatly accelerate
the convergence toward the target CC energetics. The above relationships between the semi-stochastic
CC($P$) and CC($P$;$Q$) approaches and the deterministic CCSD, CR-CC(2,3)/CR-CC(2,4), and CCSDT/CCSDTQ
theories are also helpful when debugging the CC($P$) and CC($P$;$Q$) codes.

As far as the savings in the computational effort offered by the semi-stochastic CC($P$;$Q$) methods,
when compared to their high-level CC parents, such as CCSDT or CCSDTQ, are concerned, they were already
discussed in Refs. \onlinecite{stochastic-ccpq-prl-2017,stochastic-ccpq-molphys-2020}, so here we focus
on the information relevant to the calculations discussed in Section \ref{sec3}. There are three
main factors that contribute to these savings. First, the computational times associated with the early
stages of the CIQMC walker propagations, which are sufficient to recover the parent CCSDT
or CCSDTQ energetics to within small fractions of a millihartree when the semi-stochastic CC($P$;$Q$)
framework is employed, are very short compared to the converged CIQMC runs. They are already short
when one uses FCIQMC, and they are even shorter when one replaces FCIQMC by the CISDT-MC and
CISDTQ-MC truncations.
As further elaborated on in Section \ref{sec3}, the convergence of the semi-stochastic CC($P$;$Q$)
calculations toward the parent CCSDT and CCSDTQ energies is so fast that the underlying
CIQMC computations use much smaller walker populations than those required to converge the CIQMC
propagations. They are small when one uses FCIQMC and they become even smaller when
one relies on the truncated CISDT-MC and CISDTQ-MC approaches in the CC($P$;$Q$) runs.

Second, the CC($P$) calculations using small fractions of higher--than--doubly
excited determinants, which is how the $P$ spaces used in these calculations look like when the early
stages of the CIQMC walker propagations are considered, are much faster than the parent CC computations.
For example, when the most expensive $\langle \Phi_{ijk}^{abc} | [H,T_{3}] |\Phi \rangle$ or
$\langle \Phi_{ijk}^{abc} | [\bar{H}^{(2)},T_{3}] |\Phi \rangle$ contributions to the CCSDT equations, where
$\bar{H}^{(2)} = e^{-T_{1}-T_{2}} H e^{T_{1}+T_{2}}$, are isolated and implemented using programming methods
similar to those exploited in selected CI algorithms (rather than the usual diagrammatic techniques that
assume continuous excitation manifolds labeled by all occupied and all unoccupied orbitals), one can
accelerate their determination by a factor of up to $(D/d)^{2}$, where $D$ is the number of all triples
and $d$ is the number of triples included in the $P$ space, captured with the help of CIQMC propagations.
Other contributions to the CCSDT equations that involve $T_{3}$ or the projections on the triply excited
determinants, such as $\langle \Phi_{ij}^{ab} | [H,T_{3}] |\Phi \rangle$
and $\langle \Phi_{ijk}^{abc} | [H,T_{2}] |\Phi \rangle$,
may offer additional speedups, on the order of $(D/d)$. Our current CC($P$) codes are still in the pilot
stages, but the speedups on the order of $(D/d)$ in the determination of the most expensive
$\langle \Phi_{ijk}^{abc} | [H,T_{3}] |\Phi \rangle$ (or
$\langle \Phi_{ijk}^{abc} | [\bar{H}^{(2)},T_{3}] |\Phi \rangle$)
terms are attainable. Similar remarks apply to the CC($P$)/CC($P$;$Q$) calculations aimed at converging the
CCSDTQ energetics, where one can considerably speed up the determination of the most expensive
$\langle \Phi_{ijkl}^{abcd} | [H,T_{4}] |\Phi \rangle$ or
$\langle \Phi_{ijkl}^{abcd} | [\bar{H}^{(2)},T_{4}] |\Phi \rangle$ contributions and other terms
containing the $T_{3}$ and $T_{4}$ clusters and the projections on the triply and quadruply excited
determinants. It should also be noted that the CC($P$) calculations do not require storing the
entire $T_{3}$ and $T_{4}$ vectors. The $T_{3}^{\rm (MC)}$ and $T_{4}^{\rm (MC)}$ operators use much
smaller numbers of amplitudes than their full $T_{3}$ and $T_{4}$ counterparts.

Third, the computation of the noniterative correction $\delta(P;Q)$ is much less expensive than
a single iteration of the target CC calculation. In the case of the CC($P$;$Q$) calculations
aimed at converging the CCSDT energetics, the computational time required to determine the corresponding
correction $\delta(P;Q)$ scales no worse than $\sim$$2 n_{o}^{3} n_{u}^{4}$, which is much less than
the $n_{o}^{3} n_{u}^{5}$ scaling of each iteration of CCSDT. In the case of the CC($P$;$Q$) approach
aimed at CCSDTQ, the computational time required to determine correction $\delta(P;Q)$ scales as
$\sim$$2 n_{o}^{3} n_{u}^{4}$ in the case of the contributions due to the remaining triples and is on
the order of $n_{o}^{4} n_{u}^{5}$ in the case of the quadruples part of $\delta(P;Q)$, when the
more complete CC($P$;$Q$)$_{\rm EN}$ approach is used, or
$n_{o}^{2} n_{u}^{5}$, when the CC($P$;$Q$)$_{\rm MP}$ form of $\delta(P;Q)$ is employed. This is
all much less than
the $n_{o}^{4} n_{u}^{6}$ scaling of every CCSDTQ iteration. As mentioned above, in our current
implementation of the semi-stochastic CC($P$;$Q$) approach aimed at converging the CCSDTQ energetics,
the quadruples contribution to correction $\delta(P;Q)$ is neglected, so the computational time
required to obtain $\delta(P;Q)$ scales as $\sim$$2 n_{o}^{3} n_{u}^{4}$, at worst, which points to
the usefulness of such an approximation, especially that the convergence of the resulting
CC($P$;$Q$) energies toward CCSDTQ is, as shown in Section \ref{sec3}, very fast.

\section{Numerical Examples}
\label{sec3}

In order to demonstrate the benefits offered by the semi-stochastic CC($P$;$Q$) framework,
especially the new CC($P$;$Q$) approaches implemented in this work that replace FCIQMC
by the less expensive CISDT-MC and CISDTQ-MC propagations, we applied the FCIQMC-, CISDT-MC-,
and CISDTQ-MC-driven CC($P$;$Q$) methods aimed at converging the CCSDT and CCSDTQ energetics
to a few molecular problems, for which the parent full CCSDT and CCSDTQ results had previously been
determined or were not too difficult to be recalculated. Thus, we carried out an extensive series of the
CISDT-MC- and CISDTQ-MC-driven CC($P$;$Q$) calculations, along with the analogous computations
using FCIQMC, which was utilized in our earlier study,\cite{stochastic-ccpq-prl-2017}
to examine the ability of the semi-stochastic CC($P$;$Q$) approaches using various types
of CIQMC to recover the CCSDT energetics for the F--F bond dissociation in the fluorine molecule
(Section \ref{sec3.1}) and the automerization of cyclobutadiene (Section \ref{sec3.2}).
In order to illustrate the performance of the FCIQMC- and  CISDTQ-MC-driven CC($P$;$Q$)
methods in calculations aimed at converging the CCSDTQ energetics, we considered the
symmetric stretching of the O--H bonds in the water molecule (Section \ref{sec3.3}).
We chose bond breaking in ${\rm F}_{2}$, which is accurately described by full CCSDT,
\cite{jspp-chemphys2012,jspp-jcp2012,crccl_jcp,crccl_cpl,f2bh}
since we examined the same system in our original FCIQMC- and CCSDT-MC-driven CC($P$;$Q$)
work\cite{stochastic-ccpq-prl-2017} and in the preceding deterministic CC($P$;$Q$)-based CC(t;3)
calculations reported in Ref. \onlinecite{jspp-chemphys2012}. Our choice of the automerization of
cyclobutadiene, which is accurately described by CCSDT as well,\cite{jspp-jcp2012,balkova1994} was motivated by
similar reasons. We studied this problem, where all noniterative triples corrections to CCSD, including
CCSD(T), $\Lambda$-CCSD(T), CCSD(2)$_{\rm T}$, and CR-CC(2,3) fail,\cite{jspp-jcp2012,balkova1994,tailored3}
using the deterministic CC(t;3) approach exploiting the CC($P$;$Q$) ideas in Ref.
\onlinecite{jspp-jcp2012}, and we studied it again using the semi-stochastic CC($P$;$Q$)
framework utilizing FCIQMC and CCSDT-MC in Ref. \onlinecite{stochastic-ccpq-prl-2017}.
We would like to explore now what the effect of replacing FCIQMC propagations by their
less expensive CISDT-MC and CISDTQ-MC counterparts on the convergence of the CC($P$;$Q$)
energies toward CCSDT is. We would also like to learn if the incorporation of the previously neglected
\cite{stochastic-ccpq-prl-2017} three-body component of the deexcitation operator ${\Lambda}^{(P)}$,
which is used to construct amplitudes $\ell_{K}(P)$ entering Eq. (\ref{mmcc-gen-delta}),
helps the accuracy of the resulting semi-stochastic CC($P$;$Q$) energies. We studied
the $C_{2v}$-symmetric double dissociation of ${\rm H_{2}O}$, since by simultaneously stretching
both O--H bonds by factors exceeding 2, one ends up with a catastrophic failure of CCSDT.
\cite{nbjspp-molphys2017,crccl_jcp,olsen-h2o} One needs an accurate description of the $T_{3}$
and $T_{4}$ clusters to obtain a more reliable description of the water potential
energy surface in that region.\cite{nbjspp-molphys2017}

Following our earlier semi-stochastic and deterministic CC($P$;$Q$) work,
\cite{stochastic-ccpq-prl-2017,jspp-chemphys2012,jspp-jcp2012,nbjspp-molphys2017} which also provides the
parent CCSDT\cite{stochastic-ccpq-prl-2017,jspp-chemphys2012,jspp-jcp2012,nbjspp-molphys2017}
and CCSDTQ\cite{nbjspp-molphys2017} energetics, we used the
cc-pVDZ,\cite{ccpvnz} cc-pVTZ,\cite{ccpvnz} and aug-cc-pVTZ \cite{augccpvnz} basis sets for
${\rm F}_{2}$ and the cc-pVDZ bases for cyclobutadiene and water. For consistency with Refs.
\onlinecite{stochastic-ccpq-prl-2017,jspp-chemphys2012,jspp-jcp2012}, in all of the
post-RHF computations for the F--F bond breaking in ${\rm F}_{2}$ and the automerization
of cyclobutadiene, the core electrons corresponding to the 1s shells of the fluorine
and carbon atoms were kept frozen. As in Refs. \onlinecite{nbjspp-molphys2017,olsen-h2o},
which provide the reference CCSDTQ data and, in the case of Ref. \onlinecite{olsen-h2o},
the geometries of the equilibrium and stretched water molecule used in our semi-stochastic CC($P$;$Q$)
calculations aimed at converging the CCSDTQ energetics, we correlated all electrons.
Each of the relevant $i$-FCIQMC (all systems),
$i$-CISDT-MC (${\rm F}_{2}$ and cyclobutadiene), and $i$-CISDTQ-MC (all systems) runs was
initiated by placing 100 walkers on the RHF reference determinant and we set the initiator
parameter $n_{a}$ at 3. All of the $i$-FCIQMC, $i$-CISDT-MC, and $i$-CISDTQ-MC propagations
used the time step $\delta \tau$ of 0.0001 a.u.

\subsection{Bond Breaking in \mbox{\boldmath${\bf F_{2}}$}}
\label{sec3.1}

\begin{figure*}
\includegraphics[scale=0.38]{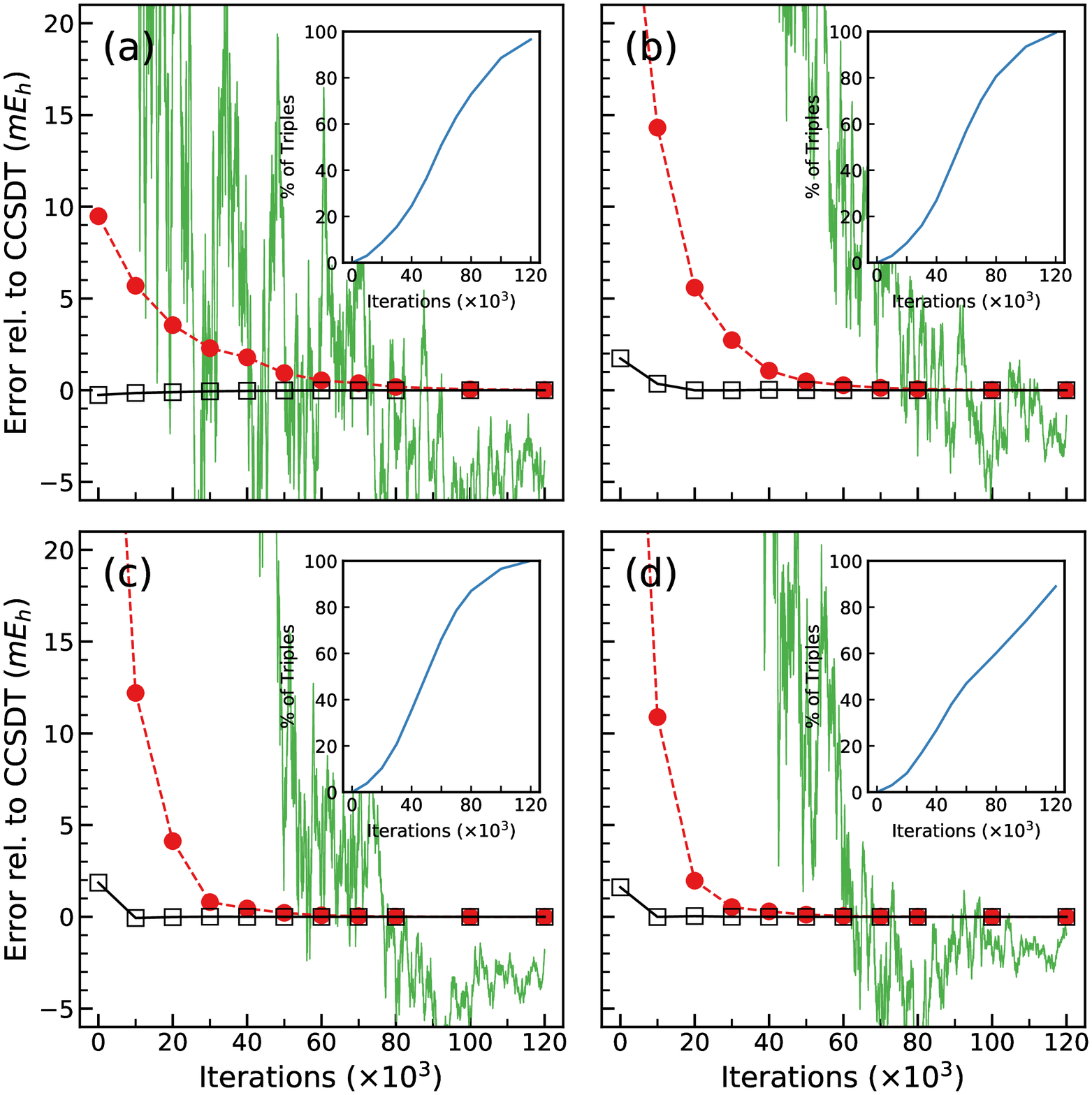}
\vspace*{-0.5em}
\caption{
Convergence of the CC($P$) (red filled circles and dashed lines)
and CC($P$;$Q$)$_{\rm EN}$ (black open squares and solid lines) energies
toward CCSDT for the ${\rm F}_{2}$/cc-pVDZ molecule in which the F--F distance $R$
was set at (a) $R_{e}$, (b) $1.5 R_{e}$, (c) $2 R_{e}$, and (d) $5 R_{e}$,
where $R_{e} = 2.66816$ bohr is the equilibrium geometry.
The $P$ spaces consisted of all singles
and doubles and subsets of triples identified during the $i$-FCIQMC
propagations with $\delta\tau = 0.0001$ a.u. (depicted by the green lines
representing the corresponding projected energies). The
$Q$ spaces consisted of the triples not captured by $i$-FCIQMC.
All energies are errors relative to CCSDT in millihartree and
the insets show the percentages of triples captured during
the $i$-FCIQMC propagations.
}
\label{figure1}
\end{figure*}

\begin{figure*}
\includegraphics[scale=0.38]{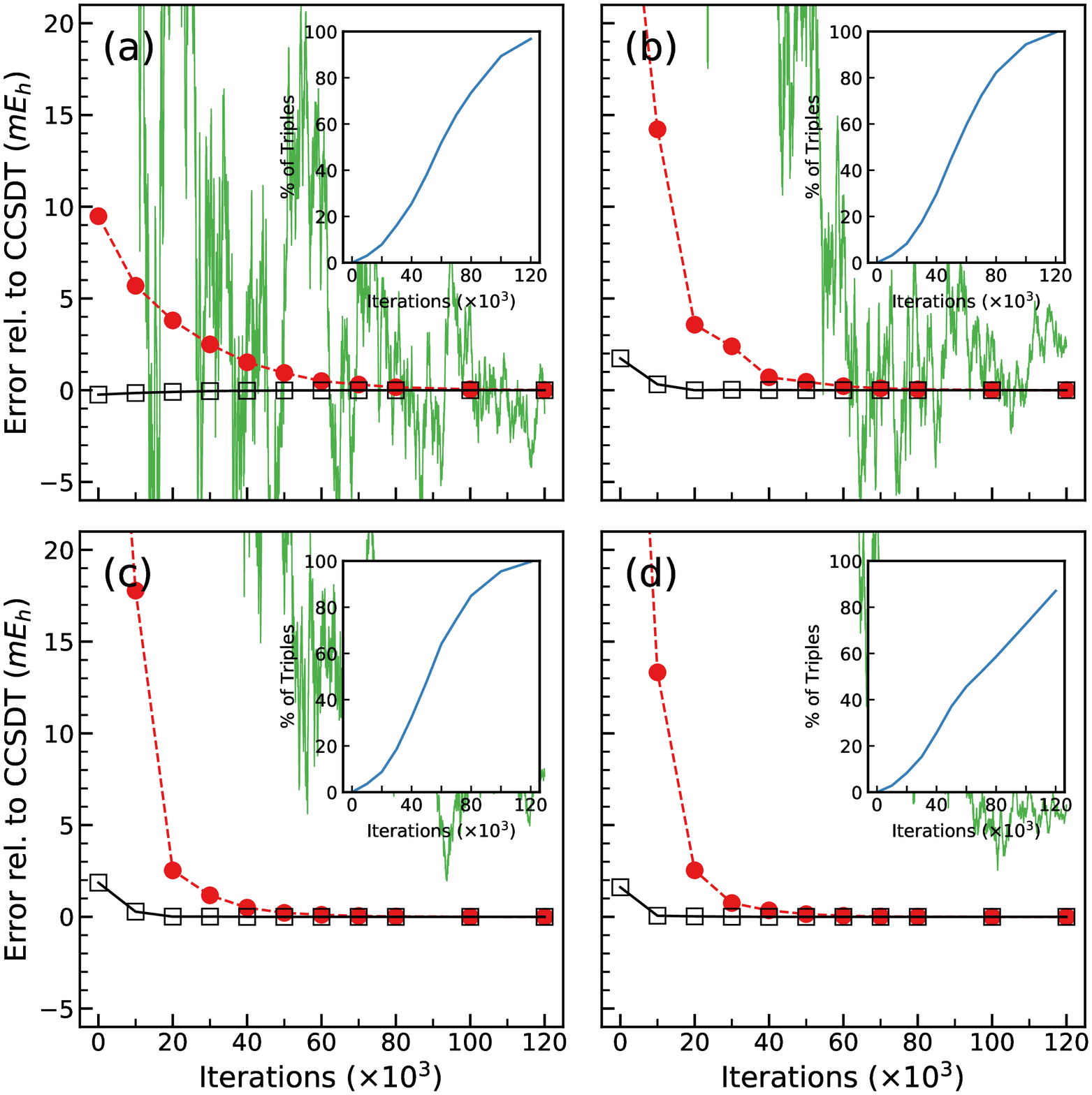}
\vspace*{-0.5em}
\caption{
Same as Fig. \ref{figure1} except that the subsets of triples
included in the CC($P$) calculations are now identified by the
$i$-CISDTQ-MC simulations and
the corresponding $Q$ spaces consist of the triples not
captured by $i$-CISDTQ-MC.
As in Fig. \ref{figure1}, the F--F distance $R$
was set at (a) $R_{e}$, (b) $1.5 R_{e}$, (c) $2 R_{e}$, and (d) $5 R_{e}$,
where $R_{e} = 2.66816$ bohr is the equilibrium geometry.
}
\label{figure2}
\end{figure*}

\begin{figure*}
\includegraphics[scale=0.38]{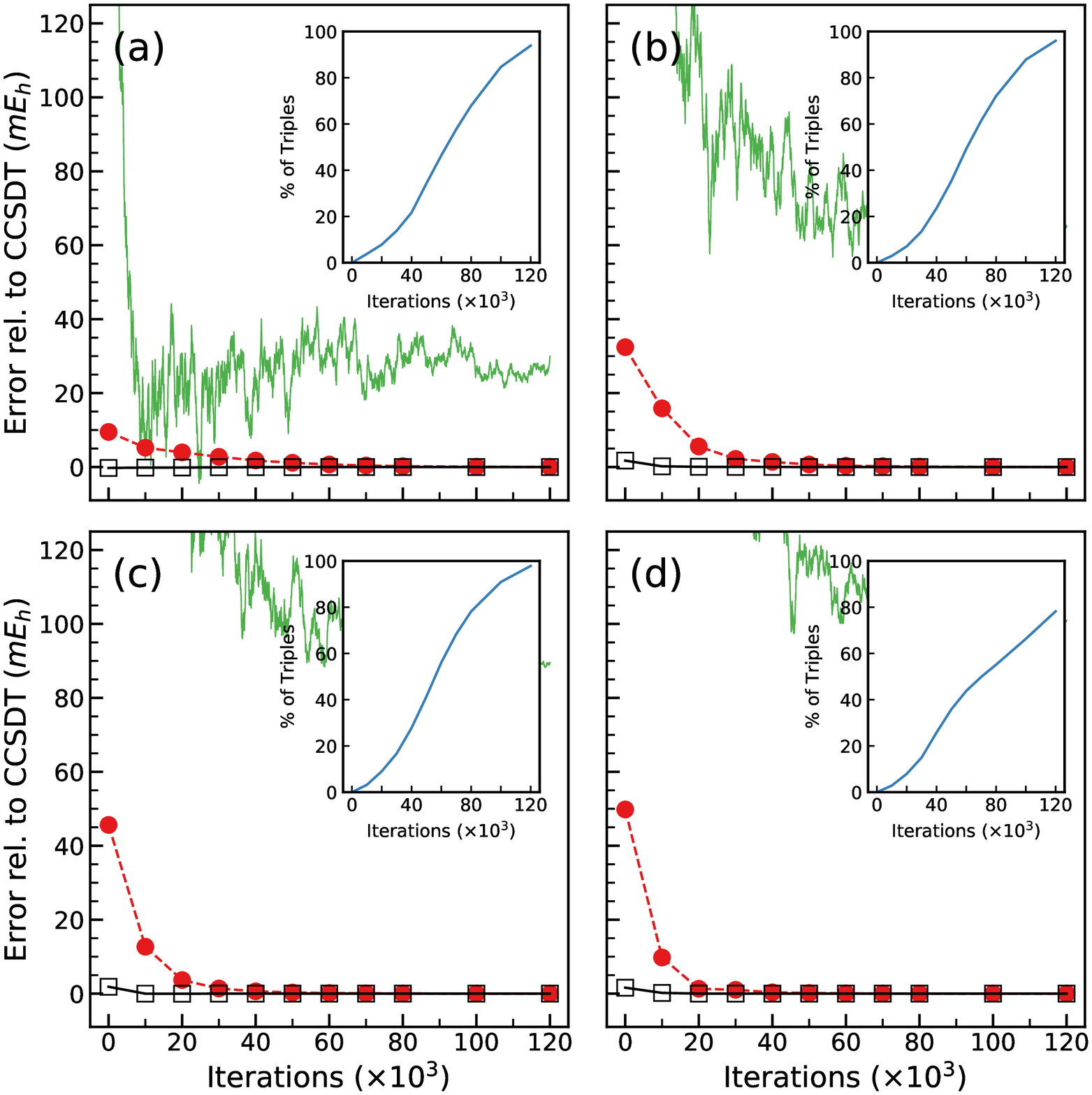}
\vspace*{-0.5em}
\caption{
Same as Fig. \ref{figure1} except that the subsets of triples
included in the CC($P$) calculations are now identified by the
$i$-CISDT-MC simulations and
the corresponding $Q$ spaces consist of the triples not
captured by $i$-CISDT-MC.
As in Fig. \ref{figure1}, the F--F distance $R$
was set at (a) $R_{e}$, (b) $1.5 R_{e}$, (c) $2 R_{e}$, and (d) $5 R_{e}$,
where $R_{e} = 2.66816$ bohr is the equilibrium geometry.
}
\label{figure3}
\end{figure*}

We begin our discussion of the semi-stochastic CC($P$;$Q$) calculations carried out in this study
with the F--F bond dissociation in the fluorine molecule, as described by the cc-pVDZ basis set
using the Cartesian components of $d$ orbitals
(see Table \ref{table1} and Figs. \ref{figure1}--\ref{figure3}). In analogy to Ref.
\onlinecite{stochastic-ccpq-prl-2017}, where our initial FCIQMC- and CCSDT-MC-based CC($P$;$Q$)
results for ${\rm F}_{2}$ were presented, we considered the equilibrium geometry $R_{e} = 2.66816$ bohr,
where the many-electron correlation effects have a predominantly dynamical character, and
three stretches of the F--F bond length $R$, including $R = 1.5 R_{e}$, $2R_{e}$, and $5R_{e}$,
which are characterized by the increasingly large nondynamical correlations. The
increasingly important role of nondynamical correlation effects as the F--F bond is stretched
is reflected in the magnitude of $T_{3}$ contributions, defined by forming the difference
of the CCSDT and CCSD energies, which grows, in absolute value, from 9.485 millihartree at $R = R_{e}$ to
32.424, 45.638, and 49.816 millihartree at $R = 1.5 R_{e}$, $2R_{e}$, and $5R_{e}$,
respectively, when the cc-pVDZ basis set is employed. The $T_{3}$ effects in the
$R = 2R_{e} - 5R_{e}$ region are so large that they exceed the depth of the CCSDT
potential well, estimated at
about 44 millihartree when the difference between the
CCSDT energies at $R = 5R_{e}$, where ${\rm F}_{2}$ is essentially dissociated, and
$R = R_{e}$ is considered. They grow with $R$ so fast that the popular perturbative CCSD(T)
correction to CCSD fails at larger F--F separations, producing the $-5.711$, $-23.596$,
and $-39.348$ millihartree errors relative to CCSDT at $R = 1.5 R_{e}$, $2R_{e}$, and $5R_{e}$,
respectively, misrepresenting the physics of $T_{3}$ correlations in the stretched
${\rm F}_{2}$ molecule.

The triples corrections to CCSD that rely on the biorthogonal moment expansions of the
CC($P$;$Q$) type, including CR-CC(2,3), work much better than CCSD(T). This is especially true when the most
complete variant of the CR-CC(2,3) approach using the Epstein--Nesbet form of the $D_{K}(P)$
denominator in determining the $\ell_{K}(P)$ amplitudes that enter the corresponding
triples correction to CCSD, abbreviated sometimes as CR-CC(2,3),D or CR-CC(2,3)$_{\rm D}$
\cite{jspp-jctc2012,nbjspp-molphys2017,crccl_jpc,ptcp2007,crccl_ijqc2} and represented in Table
\ref{table1} by the $\tau = 0$ CC($P$;$Q$)$_{\rm EN}$ results, is considered. Indeed, the
CR-CC(2,3)$_{\rm D}$ calculations reduce large errors in the CCSD(T) energies at $R = 1.5 R_{e}$,
$2R_{e}$, and $5R_{e}$ to 1.735, 1.862, and 1.613 millihartree, respectively, improving the
CCSD(2)$_{\rm T}$ or the equivalent
\cite{jspp-jctc2012,nbjspp-molphys2017,crccl_jpc,ptcp2007,crccl_ijqc2}
CR-CC(2,3),A or CR-CC(2,3)$_{\rm A}$
calculations, which adopt the  M{\o}ller--Plesset $D_{K}(P)$ denominators, at the same time
(see the $\tau = 0$ CC($P$;$Q$)$_{\rm MP}$ values in Table \ref{table1}). The CR-CC(2,3)$_{\rm D}$
approach eliminates the failure of CCSD(T) at stretched nuclear geometries, while being more
effective in capturing the physics of $T_{3}$ correlations than CCSD(2)$_{\rm T}$, but the
only way to obtain further improvements toward CCSDT is by incorporating at least some triples
in the iterative part of the calculations, relaxing the $T_{1}$ and $T_{2}$ amplitudes,
which in CCSD(T), CCSD(2)$_{\rm T}$, and CR-CC(2,3) are fixed at their CCSD values,
in the presence of the leading $T_{3}$ contributions, and correcting the resulting energies
for the remaining $T_{3}$ effects accordingly. One can do this deterministically
by turning to the previously mentioned CC(t;3) method, which uses the CC($P$;$Q$)
formalism to correct the energies obtained in the active-space CCSDt calculations for the
remaining $T_{3}$ correlation effects that the CCSDt approach did not capture,
\cite{jspp-chemphys2012,jspp-jcp2012}
or by the approximation to CC(t;3) that replaces the CC($P$;$Q$) triples correction to CCSDt
by its perturbative CCSD(T) analog, abbreviated as CCSD(T)-h.\cite{h1,h2,h3}
Alternatively, one can resort to the semi-stochastic CC($P$;$Q$)
framework advocated in this work, in which the same goal is accomplished by using full or
truncated CIQMC propagations to identify the leading triply excited determinants for the
inclusion in the underlying $P$ space without having to use active orbitals.

The semi-stochastic CC($P$;$Q$) results and the underlying CC($P$) energies shown in Table
\ref{table1} and Figs. \ref{figure1}--\ref{figure3} confirm the above expectations. Indeed,
with only about 30--40 \% of the triples in the $P$ space, captured after the relatively
short FCIQMC, CISDT-MC, and CISDTQ-MC runs at $R = R_{e}$ and $1.5 R_{e}$, and even less than
that ($\sim$15--20 \%) when the $R = 2R_{e}$ and $5R_{e}$ geometries are considered,
the errors in the uncorrected CC($P$) energies relative to their CCSDT parents are already on
the order of 1 millihartree or smaller. This is a massive error reduction compared to
the initial, $\tau = 0$, CC($P$), i.e., CCSD energy values, especially at the larger F--F separations,
where the differences between the CCSD and CCSDT energies are as high as 45.638 millihartree
at $R = 2R_{e}$ or 49.816 millihartree at $R = 5R_{e}$. The CC($P$;$Q$) corrections based on
Eq. (\ref{mmcc-gen-delta}) accelerate the
convergence toward CCSDT even further, allowing one to reach the submillihartree accuracy levels
relative to the parent CCSDT energetics almost instantaneously, out of the early stages
of the FCIQMC, CISDT-MC, and CISDTQ-MC propagations, when no more than 10 \% of all triples
are included in the corresponding $P$ spaces
and when the total numbers of walkers used in the CIQMC runs represent tiny fractions of
the walker populations required to converge these runs.

The CC($P$;$Q$)$_{\rm EN}$ correction,
which adopts the Epstein--Nesbet form of the $D_{K}(P)$ denominator in determining the
$\ell_{K}(P)$ amplitudes entering Eq. (\ref{mmcc-gen-delta}), is particularly effective
in this regard. With less than 10 \% triples in the stochastically determined $P$ spaces,
captured after 20,000 or fewer $\delta \tau = 0.0001$ a.u. MC iterations, where, as shown in
Figs. \ref{figure1}--\ref{figure3}, the FCIQMC,
CISDT-MC, and CISDTQ-MC runs are very far from convergence, the differences between the
CC($P$;$Q$)$_{\rm EN}$ energies and their CCSDT parents are on the order of 0.1 millihartree,
being usually even smaller. This is not only true at the equilibrium geometry, but also at
the larger values of $R$, including $R = 5R_{e}$, where the F--F bond in ${\rm F}_{2}$ is already
{\it de facto} broken.
As shown in Table S.1 of the supplementary material, the total numbers of walkers corresponding to
20,000 $\delta \tau = 0.0001$ a.u. MC iterations initiated by placing 100 walkers on the RHF reference
determinant $|\Phi\rangle$, which range from about 6,300 to 9,400 when one uses FCIQMC, 5,800 to 7,600 when FCIQMC
is replaced by CISDTQ-MC, and 3,500 to 4,300 when the CISDT-MC approach is employed, represent tiny
fractions of the walker populations at $\tau = 12.0$ a.u., where we stopped our CIQMC runs
(0.02--0.53 \% in the case of FCIQMC, 0.07--0.72 \% in the case of CISDTQ-MC, and
0.21--1.78 \% in the CISDT-MC case, where total walker populations are smallest).
When we perform somewhat longer FCIQMC, CISDT-MC, and CISDTQ-MC propagations,
allowing them to capture about 40--50 \% of the triples in the $P$ space, when $R = R_{e}$ and
$1.5R_{e}$, and 20--30 \% when $R \geq 2 R_{e}$, the CC($P$;$Q$)$_{\rm EN}$ calculations
recover the CCSDT energetics to within 10 or so microhartree.
This happens after 50,000--60,000 $\delta \tau = 0.0001$ a.u. MC iterations, when $R = R_{e}$ and $1.5R_{e}$,
and 30,000--40,000 MC time steps when $R \geq 2 R_{e}$, i.e., when the underlying
CIQMC propagations are still in their early stages (cf. Figs. \ref{figure1}--\ref{figure3}).
As demonstrated in Table S.1 of the supplementary material, even in this case
the total numbers of walkers characterizing the FCIQMC, CISDT-MC, and CISDTQ-MC runs
used to obtain these highly accurate CC($P$;$Q$)$_{\rm EN}$ results remain much smaller
than the walker populations required to converge the CIQMC runs. In the case of FCIQMC,
they are about 60,000 or 1--5 \% of the walker populations at $\tau = 12.0$ a.u., where we stopped
our CIQMC propagations, for $R = R_{e}$ and $1.5R_{e}$ and about 20,000--50,000 or 0.1--0.2 \% of the
walker populations at $\tau = 12.0$ a.u. when the $R \geq 2 R_{e}$ region is explored. They are even
smaller when the truncated CIQMC approaches, especially CISDT-MC, are utilized. In the case of CISDT-MC,
the total numbers of walkers allowing us to converge the CCSDT energetics using the semi-stochastic
CC($P$;$Q$)$_{\rm EN}$ calculations to within $\sim$10 microhartree are as little as $\sim$20,000
or 5--12 \% of the total walker populations at $\tau = 12.0$ a.u., which themselves are 6--12 times
smaller than those used at $\tau = 12.0$ a.u. by FCIQMC, for $R = R_{e}$ and $1.5R_{e}$ and about 10,000
or 1 \% of the CISDT-MC walker populations at $\tau = 12.0$ a.u., which themselves are 4--5 \% of their
$\tau = 12.0$ a.u. FCIQMC counterparts, when $R \geq 2 R_{e}$.
The CC($P$;$Q$)$_{\rm MP}$
correction, in which the Epstein--Nesbet $D_{K}(P)$ denominator, Eq. (\ref{eq:denom}), in
the definition of $\ell_{K}(P)$ amplitudes entering Eq. (\ref{mmcc-gen-delta}) is replaced by
its simplified M{\o}ller--Plesset form, is not as accurate as CC($P$;$Q$)$_{\rm EN}$, but
it still accelerates the convergence of the underlying CC($P$) energies, allowing
one to recover the parent CCSDT energies to within $\sim$0.1 millihartree once about 40 \%
($R = R_{e}$ and $1.5 R_{e}$) or 15--20 \% ($R = 2R_{e}$, and $5R_{e}$) of the triples are captured
by the FCIQMC, CISDT-MC, and CISDTQ-MC propagations.

The results shown in Table \ref{table1} and Figs. \ref{figure1}--\ref{figure3} demonstrate
that it is practically irrelevant whether one uses FCIQMC or one of its less expensive truncated
forms, such as CISDT-MC and CISDTQ-MC examined in this study, to identify the leading triply
excited determinants for the inclusion in the $P$ space used in the CC($P$;$Q$) and the underlying
CC($P$) calculations. Clearly, as $\tau$ approaches $\infty$, the FCIQMC, CISDT-MC, and CISDTQ-MC
propagations converge to completely different limits (FCI in the case of FCIQMC, CISDT-MC in the
case of CISDT-MC, and CISDTQ in the case of CISDTQ-MC), but this has virtually no impact on the
convergence patterns observed in our semi-stochastic CC($P$) and CC($P$;$Q$) calculations.
This is a consequence of the fact that the uncorrected CC($P$) and corrected CC($P$;$Q$)
computations are capable of recovering the parent high-level CC energetics, such as those
corresponding to full CCSDT discussed
in this subsection, based on the information extracted from the early stages of the corresponding
CIQMC runs. In particular, if we are targeting CCSDT, all we need from the CIQMC calculations is a meaningful
list of the leading triply excited determinants, which any CIQMC calculation that is allowed to sample the triples
subspace of the Hilbert space, even the crude CISDT-MC approach, can provide. One can see, for example, in Table
\ref{table1} that the fractions of triples captured by the FCIQMC, CISDT-MC, and CISDTQ-MC runs
at the various numbers of MC iterations (various propagation times $\tau$) are very similar. Detailed
inspection of the corresponding lists of triply excited determinants shows that while the numbers of
walkers on the individual determinants may substantially differ, the lists of triples identified by
the FCIQMC, CISDT-MC, and CISDTQ-MC propagations, especially the more important ones that result in
larger $T_{3}^{\rm (MC)}$ amplitudes in the subsequent deterministic CC($P$) steps,
are not much different. Once the lists of the leading triples are identified, we turn to the
CC($P$) computations, correcting them for the remaining triples not captured by CIQMC,
and this makes the semi-stochastic CC($P$) and CC($P$;$Q$) calculations rather insensitive
to the type of the CIQMC approach used to construct these lists.

All of the above observations regarding the ability of the semi-stochastic CC($P$;$Q$)
calculations using the FCIQMC, CISDT-MC, and CISDTQ-MC propagations to rapidly converge
the full CCSDT energetics remain true when the cc-pVDZ basis set is replaced by its larger
cc-pVTZ and aug-cc-pVTZ counterparts (both using the spherical components of $d$ and $f$ functions).
This is illustrated in Table \ref{table2}, where
we examine the stretched ${\rm F}_{2}$ molecule, in which the F--F distance $R$ is set at $2R_{e}$.
We chose $R = 2R_{e}$, since, in analogy to the previously discussed cc-pVDZ basis set,
the $T_{3}$ effects at this geometry, obtained by calculating differences of the respective
CCSDT and CCSD energies, which are $-62.819$ millihartree, when the cc-pVTZ basis set is
employed, and $-65.036$ millihartree, when the aug-cc-pVTZ basis is used, are not only very
large, but also larger, in absolute value, than the corresponding CCSDT dissociation energies
(differences between the CCSDT energies at $R = 5R_{e}$, where the F--F bond is broken, and $R = R_{e}$ 
obtained with the cc-pVTZ and aug-cc-pVTZ basis sets are
about 57 and 60 millihartree, respectively). We also chose it,
since the $R = 2R_{e}$ stretch of the F--F bond length is large enough for the conventional CCSD(T)
approach to fail in a major way when the cc-pVTZ and aug-cc-pVTZ basis sets are employed,
resulting in the $-26.354$ and $-27.209$ millihartree errors relative to CCSDT, respectively.
The CCSD(2)$_{\rm T}$ correction to CCSD or the equivalent CR-CC(2,3)$_{\rm A}$ approximation,
represented in Table \ref{table2} by the $\tau = 0$ CC($P$;$Q$)$_{\rm MP}$ results, helps,
but large differences between the CCSD(2)$_{\rm T}$ and CCSDT energies, of 9.211 millihartree in the
cc-pVTZ case and 9.808 millihartree when the aug-cc-pVTZ basis set is employed, remain.
The CR-CC(2,3)$_{\rm D}$ approach, represented in Table \ref{table2} by the
$\tau = 0$ CC($P$;$Q$)$_{\rm EN}$ data, is more effective than other triples corrections
to CCSD, reducing the large errors relative to CCSDT observed in the CCSD(T) and CCSD(2)$_{\rm T}$
calculations to 4.254 (cc-pVTZ) and 5.595 (aug-cc-pVTZ) millihartree,
but none of the above results are as good as the energies resulting from the semi-stochastic
CC($P$;$Q$) calculations using FCIQMC, CISDT-MC, and CISDTQ-MC.

Indeed, as shown in Table \ref{table2}, we observe a rapid error reduction relative to the parent CCSDT
data once we start migrating the triply excited determinants identified during the FCIQMC, CISDT-MC, 
and CISDTQ-MC propagations into the underlying $P$ space. With about 20--30 \% (cc-pVTZ)
or 30--40 \% (aug-cc-pVTZ) of the triples in the $P$ space, the 62.819 and 65.036 millihartree
errors resulting from the initial CCSD ($\tau = 0$ CC($P$)) computations decrease to a 1--2
millihartree level when the CC($P$) method is employed.
The CC($P$;$Q$) corrections due to the remaining triples not captured by
FCIQMC, CISDT-MC, and CISDTQ-MC accelerate the convergence toward CCSDT even further, with
the semi-stochastic CC($P$;$Q$)$_{\rm EN}$ approach being particularly efficient in this regard.
With only 2--4 \% of the triples in the stochastically determined $P$ spaces, captured after
20,000--30,000 $\delta \tau = 0.0001$ a.u. MC iterations, which are the very early stages of the
FCIQMC, CISDT-MC, and CISDTQ-MC propagations, the CC($P$;$Q$)$_{\rm EN}$ calculations recover
the full CCSDT energetics corresponding to the cc-pVTZ and aug-cc-pVTZ basis sets to within
0.1--0.2 millihartree. After 50,000 (cc-pVTZ) or 60,000 (aug-cc-pVTZ) MC iterations,
where the FCIQMC, CISDT-MC, and CISDTQ-MC runs are still far from convergence, capturing
only about 20--30 \% (cc-pVTZ) or 30--40 \% (aug-cc-pVTZ) of the triples, the errors in
the CC($P$;$Q$)$_{\rm EN}$ energies relative to CCSDT reduce to a 10 microhartree level.
Similarly to the previously discussed calculations using the cc-pVDZ basis set,
the total numbers of walkers characterizing the CIQMC propagations
that allowed us to reproduce the CCSDT/cc-pVTZ and CCSDT/aug-cc-pVTZ energies
so accurately represent tiny fractions of the walker populations required to converge the  CIQMC runs
(see Table S.2 of the supplementary material). For example, the total number of walkers corresponding
to 30,000 $\delta \tau = 0.0001$ a.u. FCIQMC iterations initiated by placing 100 walkers on the
RHF reference determinant, which enable the FCIQMC-driven CC($P$;$Q$)$_{\rm EN}$ approach to recover
the CCSDT/aug-cc-pVTZ energy of ${\rm F}_{2}$ at $R = 2R_{e}$ to within $\sim$0.1 millihartree,
is only about 200,000. As shown in Table S.2 of the supplementary material,
this translates into less than 0.1 \% of the total walker population used
by the FCIQMC run at $\tau = 10.0$ a.u., where we terminated our CIQMC propagations.
With about 2 million walkers in the FCIQMC computation,
reached after 50,000 $\delta \tau = 0.0001$ a.u. MC iterations, i.e.,
with less than 1 \% of the total walker population at $\tau = 10.0$ a.u., the difference between the FCIQMC-based
CC($P$;$Q$)$_{\rm EN}$ energy and its CCSDT parent reduces to 16 microhartree. The analogous
$\tau = 5.0$ a.u. CISDTQ-MC and CISDT-MC calculations, which allow the CC($P$;$Q$)$_{\rm EN}$
approach to recover the CCSDT/aug-cc-pVTZ energy of ${\rm F}_{2}$ at $R = 2R_{e}$ to within
19 and 38 microhartree, respectively, use even smaller numbers of
walkers, namely, a little over 1 million in the case of CISDTQ-MC and less than
300,000 in the CISDT-MC case.
In analogy to the cc-pVDZ basis set, the CC($P$;$Q$)$_{\rm MP}$ correction is
less accurate than its CC($P$;$Q$)$_{\rm EN}$ counterpart when the cc-pVTZ and aug-cc-pVTZ
basis sets are employed, recovering the CCSDT energetics to within 0.1--0.2 millihartree
after 50,000 rather than 20,000--30,000 MC iterations, i.e., after about 20--30 \% rather
than 2--4 \% of the triples are captured by the CIQMC propagations, but the overall
error reduction compared to the underlying CC($P$) calculations or the various noniterative
triples corrections to CCSD is still impressive. 

Similarly to the cc-pVDZ basis set, the semi-stochastic CC($P$;$Q$) calculations using larger
cc-pVTZ and aug-cc-pVTZ bases are rather insensitive to the type of the CIQMC approach used
to identify the leading triples for the inclusion in the $P$ space. Based on the results in
Table \ref{table2}, one might try to argue that the
energies obtained with the uncorrected CC($P$) approach using the CISDT-MC propagations are
characterized by slower convergence compared to their CISDTQ-MC- and FCIQMC-driven counterparts,
but this would be misleading, since CISDT-MC captures the leading triples at a somewhat slower
rate, while being less expensive than CISDTQ-MC and FCIQMC at the same time. For example,
the CISDT-MC-driven CC($P$) computations for ${\rm F}_{2}$ at $R = 2 R_{e}$ using the
cc-pVTZ basis set need 60,000 $\delta \tau = 0.0001$ a.u.
MC iterations to reach a $\sim$1 millihartree accuracy relative to the corresponding
CCSDT energy. The CC($P$) approach using CISDTQ-MC and FCIQMC reaches the same accuracy
level sooner, after 50,000 MC iterations. One should keep in mind, however,
that it takes 60,000 MC time steps for the CISDT-MC propagation to capture about
30 \% of the triples, needed to reach a $\sim$1 millihartree accuracy level in the
subsequent CC($P$) calculations, and the analogous CISDTQ-MC and FCIQMC runs capture a
similar fraction of the triples after 50,000 time steps. Ultimately, one needs
to remember that all CIQMC-driven CC($P$) computations considered in this subsection
converge to CCSDT as $\tau \rightarrow \infty$, independent of the type of the CIQMC
approach used to define the underlying $P$ spaces, as long as the CIQMC propagation
is allowed to spawn walkers on the triply excited determinants. Perhaps more importantly,
the CC($P$;$Q$) corrections to the CC($P$) energies make the convergence toward CCSDT
not only much faster, but also less dependent on the type of the CIQMC approach used
in the calculations, since they take care of the triples that were not captured by the
respective QMC propagations.

Before discussing our next molecular example, it is worth pointing out that the FCIQMC-driven
CC($P$;$Q$) calculations reported in Tables \ref{table1} and \ref{table2} and Fig. \ref{figure1},
in which, as explained in Section \ref{sec2.2}, we used complete representations of $\bar{H}^{(P)}$
and ${\Lambda}^{(P)}$ in determining corrections $\delta(P;Q)$, approach the parent CCSDT energetics of
the stretched ${\rm F}_{2}$ system in the early stages of the underlying FCIQMC propagations
faster than the analogous calculations reported in Ref. \onlinecite{stochastic-ccpq-prl-2017},
where the three-body component of ${\Lambda}^{(P)}$ was neglected. For example, the CC($P$;$Q$)
energies of ${\rm F}_{2}$ at $R = 2 R_{e}$ using the aug-cc-pVTZ basis set obtained in this
work after 10,000, 20,000, and 30,000 $\delta \tau = 0.0001$ a.u. MC iterations of the underlying
FCIQMC propagation differ from the corresponding CCSDT energy by 1.594, 0.382, and 0.138 millihartree,
respectively (see Table \ref{table2}). The analogous energy differences reported in Ref.
\onlinecite{stochastic-ccpq-prl-2017}, of 3.770, 1.661, and 0.454 millihartree, respectively, are
noticeably larger (see Table II in the Supplemental Material to Ref. \onlinecite{stochastic-ccpq-prl-2017}).
In fact, by comparing the FCIQMC-, CISDT-MC-, and CISDTQ-MC-based CC($P$;$Q$) energies shown in
Tables \ref{table1} and \ref{table2} and Figs. \ref{figure1}--\ref{figure3}, determined by treating
the deexcitation operator ${\Lambda}^{(P)}$ in Eq. (\ref{eq:ell}) fully, i.e., by defining
${\Lambda}^{(P)}$ as $\Lambda_{1} + \Lambda_{2} + \Lambda_{3}^{\rm (MC)}$, with their FCIQMC-
and CCSDT-MC-based counterparts obtained in Ref. \onlinecite{stochastic-ccpq-prl-2017}, where
${\Lambda}^{(P)}$ was approximated by $\Lambda_{1} + \Lambda_{2}$, we can conclude that as long as
$\Lambda_{3}^{\rm (MC)}$ is not neglected one can replace FCIQMC by CISDTQ-MC or, even, CISDT-MC
and still improve the rate of convergence of the CC($P$;$Q$) energies toward CCSDT in the early
stages of the QMC propagations compared to that reported in Ref. \onlinecite{stochastic-ccpq-prl-2017}.

The above observations, combined with the superior
performance of the CC($P$;$Q$)$_{\rm EN}$ approach compared to its CC($P$;$Q$)$_{\rm MP}$
counterpart, suggest that a complete treatment of correction $\delta(P;Q)$, as dictated by Eqs.
(\ref{mmcc-gen-delta}), (\ref{eq:ell}), and (\ref{eq:denom}), is more important, especially when one is
interested in accelerating convergence of the semi-stochastic CC($P$;$Q$) calculations for stretched
or more multireference molecules in the early stages of the QMC propagations, than the actual
type of the underlying CIQMC approach. It is interesting to examine if the same remains
true when other molecular examples, including those discussed in the next two subsections,
are considered.

\subsection{Automerization of Cyclobutadiene}
\label{sec3.2}

\begin{figure*}
\includegraphics[scale=0.8]{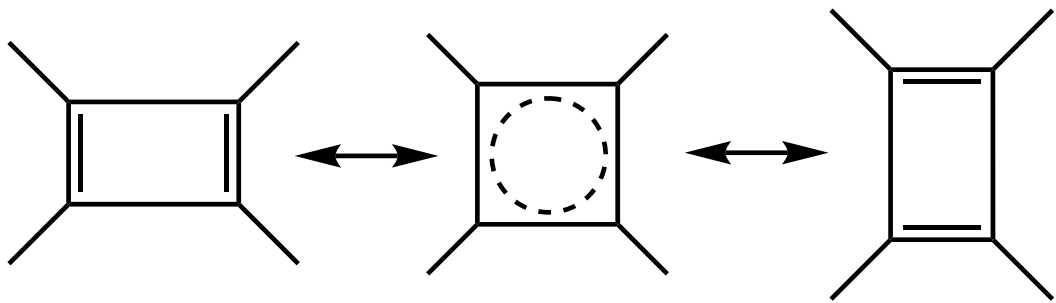}
\vspace*{-0.5em}
\caption{
The key molecular structures defining the automerization of cyclobutadiene.
The leftmost and rightmost structures represent the degenerate
reactant/product minima, whereas the structure in the center
corresponds the transition state.
}
\label{figure4}
\end{figure*}

\begin{figure*}
\includegraphics[scale=0.38]{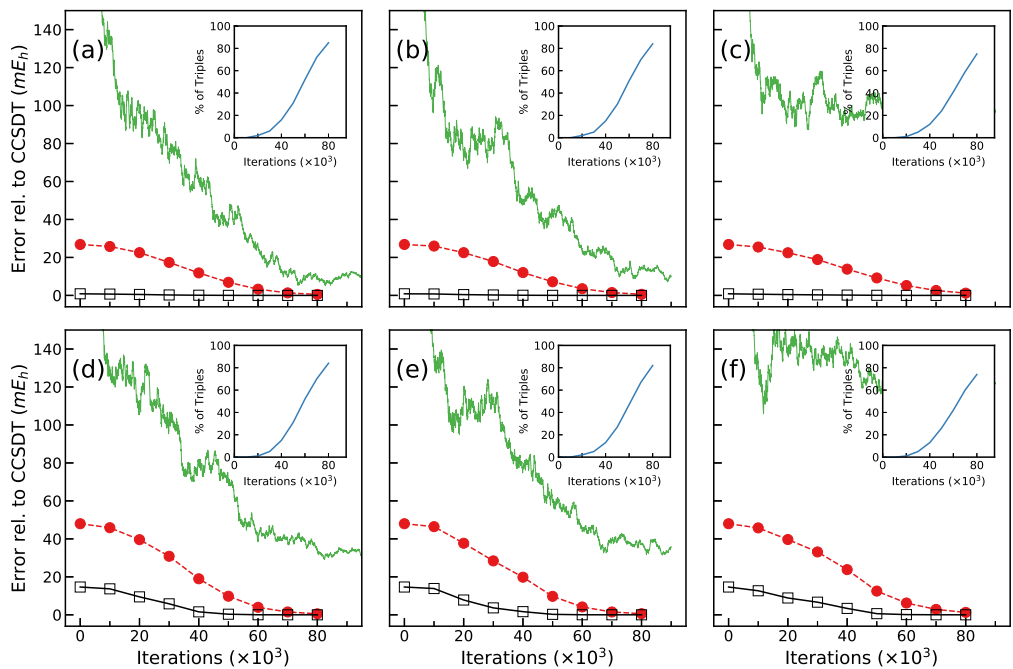}
\vspace*{-0.5em}
\caption{
Convergence of the CC($P$) (red filled circles and dashed lines) and
CC($P$;$Q$)$_{\rm EN}$ (black open squares and solid lines) energies toward
CCSDT for the reactant [panels (a)--(c)] and transition-state [panels (d)--(f)]
structures defining the automerization of cyclobutadiene, as described by the
cc-pVDZ basis set. The relevant $i$-CIQMC runs (all using $\delta\tau = 0.0001$ a.u.)
are depicted by the green lines representing the corresponding projected energies.
Panels (a) and (d) correspond to the calculations
in which the $P$ spaces employed in the CC($P$) steps consisted of all
singles and doubles and subsets of triples identified during the $i$-FCIQMC
propagations; the $Q$ spaces needed to
define the corresponding $\delta(P;Q)$ corrections consisted of the triples that
were not captured by $i$-FCIQMC. Panels (b) and (e) correspond to the calculations
in which the $P$ spaces employed in the CC($P$) steps consisted of all 
singles and doubles and subsets of triples identified during the $i$-CISDTQ-MC
propagations; in this case, the $Q$ spaces
needed to define the $\delta(P;Q)$ corrections consisted of the triples that
were not captured by $i$-CISDTQ-MC. Panels (c) and (f) correspond to the calculations
in which the $P$ spaces employed in the CC($P$) steps consisted of all
singles and doubles and subsets of triples identified during the $i$-CISDT-MC
propagations; in this case, the $Q$ spaces
needed to define the $\delta(P;Q)$ corrections consisted of the triples that
were not captured by $i$-CISDT-MC. 
All reported energies are errors relative to CCSDT in millihartree.
The insets show the percentages of triples captured during
the relevant $i$-CIQMC propagations.
}
\label{figure5}
\end{figure*}

Our next example is the challenging and frequently studied
\cite{jspp-jcp2012,balkova1994,tailored3,CBDexp,hess1983,carpenter1983,goddard1986,carsky1988,michl1993,%
wu2002,krylov2004,demel-pittner-nonit,worth2006,MR-AQCC,mkcc_our1,karadakov2008,icmrcc6,BCCC5,%
paldus-li-succ-2009,evangelista-mrsrg-2019,hfs-aci-ec-cc-2021} automerization of cyclobutadiene
(see Fig. \ref{figure4}). In this case, in order to obtain reliable energetics using computational
means, especially the activation energy, one has to provide an accurate and well-balanced description
of the nondegenerate closed-shell reactant (or the equivalent product) species,
in which the many-electron correlation effects have a predominantly dynamical character, and
the quasi-degenerate, biradicaloid transition state characterized by substantial non-dynamical
correlations. Experiment suggests that the activation energy for the automerization of cyclobutadiene
is somewhere between 1.6 and 10 kcal/mol.\cite{CBDexp,carpenter1983} The most accurate single- and
multi-reference calculations performed to date, reviewed, for example, in Refs.
\onlinecite{jspp-jcp2012,tailored3,evangelista-mrsrg-2019}, imply that the purely electronic
value of the energy barrier falls into the 6--10 kcal/mol range. In particular, as pointed out
in Ref. \onlinecite{jspp-jcp2012} (cf., also, Ref. \onlinecite{balkova1994}), one can obtain
a reliable description of the activation energy using the
full CCSDT approach. Given this information and the methodological nature of the present
study, in which we had to perform a large number of semi-stochastic CC($P$) and CC($P$;$Q$)
calculations, exploring three different types of the CIQMC method, including FCIQMC, CISDT-MC, and CISDTQ-MC,
and probing many values of the QMC propagation time $\tau$, in a discussion below we focus on converging
the CCSDT energetics obtained using the spherical cc-pVDZ basis set. As shown in Ref. \onlinecite{jspp-jcp2012}
and Table \ref{table3}, the
CCSDT/cc-pVDZ activation energy characterizing the automerization of cyclobutadiene, assuming the
reactant/product and transition-state geometries obtained with the multireference average-quadratic CC
(MR-AQCC) approach\cite{mraqcc1,mraqcc2} in Ref. \onlinecite{MR-AQCC}, which we adopt in the CC($P$)
and CC($P$;$Q$) calculations reported in this work as well, is 7.627 kcal/mol, in reasonable agreement
with the most accurate {\it ab initio} results reported to date. The results of our semi-stochastic
CC($P$) and CC($P$;$Q$) calculations, aimed at recovering the CCSDT/cc-pVDZ energetics of the reactant and
transition-state species and the corresponding activation energy using the FCIQMC, CISDT-MC, and
CISDTQ-MC propagations to identify the leading triply excited determinants for constructing the
underlying $P$ spaces, are summarized in Table \ref{table3} and Fig. \ref{figure5}.

As already mentioned, all of the noniterative triples corrections to CCSD, including CCSD(T), $\Lambda$-CCSD(T),
CCSD(2)$_{\rm T}$, and CR-CC(2,3), perform very poorly in this case, producing activation barriers in a
16--17 kcal/mol range when the cc-pVDZ basis set is considered,\cite{jspp-jcp2012,tailored3} instead of
$\sim$8 kcal/mol obtained with CCSDT (it should be noted that the 16--17 kcal/mol values are also way
outside the experimentally derived and most accurate theoretically determined ranges of 1.6--10 kcal/mol
and 6--10 kcal/mol, respectively). They improve the CCSD activation energy, which is even worse
(about 21 kcal/mol; see the $\tau = 0$ CC($P$) barrier in Table \ref{table3}), but the improvements
offered by the noniterative triples corrections to CCSD are far from sufficient.
This, in particular, applies to the ${\rm CCSD(2)}_{\rm T} = {\rm CR\mbox{-}CC(2,3)}_{\rm A}$
and CR-CC(2,3)$_{\rm D}$ approaches, represented in Table \ref{table3} by the $\tau = 0$
CC($P$;$Q$)$_{\rm MP}$ and CC($P$;$Q$)$_{\rm EN}$ data, respectively, where errors in the
resulting activation energies relative to CCSDT are 9.611 kcal/mol (126 \%) in the former case
and 8.653 kcal/mol (113 \%) in the case of the latter method. As explained in Ref.
\onlinecite{jspp-jcp2012}, the poor performance of the noniterative triples corrections to
CCSD in describing the automerization of cyclobutadiene is a consequence of neglecting the
coupling between the $T_{3}$ clusters and their lower-order $T_{1}$ and $T_{2}$ counterparts,
which is accounted for in CCSDT, but ignored in methods such as CCSD(T), $\Lambda$-CCSD(T),
CCSD(2)$_{\rm T}$, and CR-CC(2,3). This coupling is particularly large at the transition-state
geometry, where the magnitude of $T_{3}$ contributions, defined as the absolute value
of the difference between the CCSDT and CCSD energies, is nearly 48 millihartree,
when the cc-pVDZ basis set is employed, and where errors in the CCSD(T),
$\Lambda$-CCSD(T), CCSD(2)$_{\rm T}$, and CR-CC(2,3) energies relative to CCSDT
range from about 14 to 20 millihartree, as opposed to $\sim$1--5 millihartree
obtained for the reactant (see Refs. \onlinecite{jspp-jcp2012,tailored3}
and Table \ref{table3}). In analogy to bond breaking in ${\rm F}_{2}$, if we want to
capture the coupling of the $T_{1}$, $T_{2}$, and $T_{3}$ clusters without having to solve full
CCSDT equations, while preserving the idea of noniterative triples corrections to energies
obtained in lower-order CC calculations, we must solve for the $T_{1}$ and $T_{2}$ amplitudes,
which in the CCSD(T), $\Lambda$-CCSD(T), CCSD(2)$_{\rm T}$, and CR-CC(2,3) approaches
are obtained with CCSD, in the presence of the dominant $T_{3}$ components 
by incorporating some triples in the iterative CC steps, and then
correct the resulting energies for the remaining $T_{3}$ effects
neglected in the CC iterations. Again, this can be done deterministically
by solving the active-space CCSDt equations, in which the dominant $T_{3}$ amplitudes
are selected using active orbitals, and correcting the CCSDt energies for the remaining
$T_{3}$ correlations using the CC($P$;$Q$) corrections $\delta(P;Q)$, as in the
CC(t;3) calculations reported in Ref. \onlinecite{jspp-jcp2012}, or by turning
to the semi-stochastic form of the CC($P$;$Q$) formalism pursued in this study,
which eliminates the need for defining active orbitals, when identifying the leading
triples, by resorting to CIQMC propagations. Interestingly, using the CCSD(T)-type
correction to CCSDt, as in the aforementioned CCSD(T)-h approach, in the calculations
for the automerization of cyclobutadiene worsens the activation energies obtained with CCSDt,
moving them away from their parent CCSDT values.\cite{jspp-jcp2012} This underlines the
significance of treating corrections $\delta(P;Q)$ due to the correlation effects outside the
underlying $P$ spaces as completely as possible, following Eqs. (\ref{mmcc-gen-delta}), (\ref{eq:ell}),
and (\ref{eq:denom}), avoiding drastic approximations in these equations that lead to the triples
corrections of CCSD(T).

As shown in Table \ref{table3} and Fig. \ref{figure5}, the semi-stochastic CC($P$;$Q$)
calculations using FCIQMC, CISDT-MC, and CISDTQ-MC are remarkably efficient in capturing
the desired $T_{3}$ correlation effects. Independent of the type of the CIQMC approach,
they allow us to converge the CCSDT values of
the transition-state and activation energies, which are poorly described by the
noniterative triples corrections to CCSD, to within 1--2 millihartree or 1--2 kcal/mol
out of the early stages of the CIQMC propagations,
while further improving an accurate description of the reactant by methods such as CR-CC(2,3)$_{\rm D}$.
Similarly to ${\rm F}_{2}$, the performance of the CC($P$;$Q$)$_{\rm EN}$ approach,
which uses the Epstein--Nesbet form of the $D_{K}(P)$ denominator in calculating the
$\ell_{K}(P)$ amplitudes entering Eq. (\ref{mmcc-gen-delta}), is particularly impressive.
With just 5--6 \% of the triples in the stochastically determined $P$ spaces,
captured by the FCIQMC, CISDT-MC, and CISDTQ-MC
propagations after 30,000 $\delta \tau = 0.0001$ a.u. MC iterations, i.e., almost
instantaneously, the CC($P$;$Q$)$_{\rm EN}$ approach reduces the
initial 0.848 millihartree, 14.636 millihartree, and 8.653 kcal/mol
errors in the reactant, transition-state, and activation energies relative
to CCSDT obtained in the $\tau = 0$ CC($P$;$Q$)$_{\rm EN}$ or CR-CC(2,3)$_{\rm D}$
calculations by factors of 2--4, to 0.228--0.279 millihartree, 3.648--6.651 millihartree,
and 2.146--3.999 kcal/mol, respectively. After the additional 10,000 MC time steps,
which result in capturing 12--16 \% of the triples in the underlying $P$ spaces,
errors in the CC($P$;$Q$)$_{\rm EN}$ reactant, transition-state, and activation energies
relative to their CCSDT values become 0.080--0.164 millihartree, 1.556--3.367 millihartree,
and 0.919--2.011 kcal/mol, respectively. These are remarkable improvements compared to the
initial CR-CC(2,3)$_{\rm D}$ values, especially if we realize that the early stages of the
FCIQMC, CISDT-MC, and CISDTQ-MC calculations, such as 30,000--40,000 $\delta \tau = 0.0001$ a.u.
MC time steps, are all very fast,
using, as shown in Table S.3 of the supplementary material, tiny fractions of the total
walker populations at $\tau = 8.0$ a.u., where we terminated our CIQMC runs,
and 5--6\% or 12--16 \% are small
fractions of the triples that result in large speedups in the underlying
CC($P$) calculations and significant reductions in the $T_{3}$ amplitude storage
requirements. After 50,000 MC iterations, where, as shown in
Fig. \ref{figure5}, the FCIQMC, CISDT-MC, and CISDTQ-MC runs are still
far from convergence, capturing about 20--30 \% of the triples,
i.e., still relatively small fractions of all triply excited determinants,
the CC($P$;$Q$)$_{\rm EN}$ calculations recover the CCSDT values of the
reactant, transition-state, and activation energies to within 22--57 microhartree,
0.243--0.602 millihartree, and 0.138--0.343 kcal/mol, respectively, which is a
massive error reduction compared to CR-CC(2,3)$_{\rm D}$ and other noniterative
triples corrections to CCSD.
Again, as demonstrated in Table S.3 of the supplementary material,
the total numbers of walkers used by the underlying CIQMC calculations, which allowed
the semi-stochastic CC($P$;$Q$)$_{\rm EN}$ computations to converge the CCSDT energetics so tightly,
are not only small fractions of the corresponding walker populations at $\tau = 8.0$ a.u.,
where we stopped our CIQMC propagations (about 5 \% in the case of FCIQMC, 8--9 \% in the
CISDTQ-MC case, and 15--16 \% when the CISDT-MC approach was employed), but also small
in absolute values. In the case of the $\tau = 5.0$ a.u. FCIQMC and CISDTQ-MC computations corresponding to
50,000 $\delta \tau = 0.0001$ a.u. MC time steps, they are about 2.3 million and 1.8--2.0 million,
respectively. When one switches to CISDT-MC, they go down to less than half a million.
As in the case of bond breaking in ${\rm F}_{2}$,
the CC($P$;$Q$)$_{\rm MP}$ correction, which uses the M{\o}ller--Plesset
$D_{K}(P)$ denominator in Eq. (\ref{eq:ell}) instead of its more elaborate
Epstein--Nesbet form given by Eq. (\ref{eq:denom}), is less accurate than
its CC($P$;$Q$)$_{\rm EN}$ counterpart, but its ability to accelerate
convergence of the underlying CC($P$) energies and improving the results
obtained with CR-CC(2,3) and other triples corrections to CCSD is still quite
impressive. For example, with about 20--30 \% of the triples captured by
the FCIQMC, CISDT-MC, and CISDTQ-MC propagations after 50,000 MC iterations,
the differences between the CC($P$;$Q$)$_{\rm MP}$ reactant, transition-state,
and activation energies and their CCSDT counterparts, of 0.877--1.235 millihartree,
1.488--2.238 millihartree, and 0.361--0.629 kcal/mol, are much smaller than the
analogous errors relative to CCSDT resulting from the corresponding CC($P$)
calculations, which are 6.895--9.202 millihartree, 9.727--12.495 millihartree,
and 1.601--2.067 kcal/mol, respectively, although they are not as small as the
aforementioned 22--57 microhartree, 0.243--0.602 millihartree, and 0.138--0.343 kcal/mol
errors obtained using the CC($P$;$Q$)$_{\rm EN}$ correction.

In analogy to the fluorine molecule, the semi-stochastic CC($P$;$Q$) calculations
aimed at converging the CCSDT results for the automerization of cyclobutadiene are
generally insensitive to the type of the CIQMC approach used to identify the leading triples
for the inclusion in the underlying $P$ spaces. It is sufficient to resort to the least expensive
forms of the CIQMC propagations capable of capturing the triples, such as CISDT-MC or CISDTQ-MC,
to obtain the fast convergence of the CC($P$;$Q$) reactant, transition-state, and activation energies
toward their CCSDT parents observed in
Table \ref{table3} and Fig. \ref{figure5}. Treating the CC($P$;$Q$) correction
$\delta(P;Q)$ fully, following Eqs. (\ref{mmcc-gen-delta}), (\ref{eq:ell}), and (\ref{eq:denom}),
is, however, important. We have already discussed the benefits of using the Epstein--Nesbet form of
the $D_{K}(P)$ denominator, Eq. (\ref{eq:denom}), in determining the $\ell_{K}(P)$ amplitudes
entering Eq. (\ref{mmcc-gen-delta}). A complete treatment of the deexcitation operator ${\Lambda}^{(P)}$
in Eq. (\ref{eq:ell}), which in the case of the triples corrections to the CC($P$) energies considered
here means representing it as $\Lambda_{1} + \Lambda_{2} + \Lambda_{3}^{\rm (MC)}$,
is important too. One can consider an approximation in which the three-body component
$\Lambda_{3}^{\rm (MC)}$ is neglected, which is what we did in Ref. \onlinecite{stochastic-ccpq-prl-2017},
but it is generally better, especially in the earlier stages of the CIQMC propagations,
to keep all of the relevant many-body components of ${\Lambda}^{(P)}$ in calculating the
$\ell_{K}(P)$ amplitudes that enter the CC($P$;$Q$) correction $\delta(P;Q)$.
This can be illustrated by comparing the results of the FCIQMC-driven CC($P$;$Q$)
computations shown in Table \ref{table3}, where we used a complete representation of ${\Lambda}^{(P)}$,
in which the three-body component $\Lambda_{3}^{\rm (MC)}$ was included, with the analogous results
reported in Ref. \onlinecite{stochastic-ccpq-prl-2017}, where $\Lambda_{3}^{\rm (MC)}$ was neglected.
For example, the differences between the CC($P$;$Q$)$_{\rm EN}$ reactant, transition-state, and
activation energies and their CCSDT counterparts obtained in this work after 40,000 $\delta \tau = 0.0001$ a.u.
time steps of the FCIQMC propagation are 92 microhartree, 1.556 millihartree, and
0.919 kcal/mol, respectively. The analogous energy differences reported in Ref.
\onlinecite{stochastic-ccpq-prl-2017} are 0.489 millihartree, 3.235 millihartree, and 1.7 kcal/mol, respectively, i.e., they
are substantially larger. Ultimately, when the propagation time $\tau$ becomes longer, different ways of
handling the ${\Lambda}^{(P)}$ operator or different ways of defining the $D_{K}(P)$ denominator
in Eq. (\ref{eq:ell}) become less important, but if we are interested in accurately
approximating the parent CC energetics in the early stages of the underlying CIQMC propagations,
treating these quantities fully is essential.

As shown in this subsection and Section \ref{sec3.1}, using complete representations
of the ${\Lambda}^{(P)}$ and $\bar{H}^{(P)}$ operators and the Epstein--Nesbet-type denominators
$D_{K}(P)$ in determining corrections $\delta(P;Q)$ benefits the semi-stochastic CC($P$;$Q$) calculations
aimed at converging the CCSDT energetics. In Section \ref{sec3.3}, which is the final part
of our discussion of the numerical results obtained in this work, we investigate if similar applies to the
CIQMC-driven CC($P$;$Q$) computations targeting CCSDTQ.

\subsection{Double Dissociation of \mbox{\boldmath${\rm H_{2}O}$}}
\label{sec3.3}

\begin{figure*}
\includegraphics[scale=0.38]{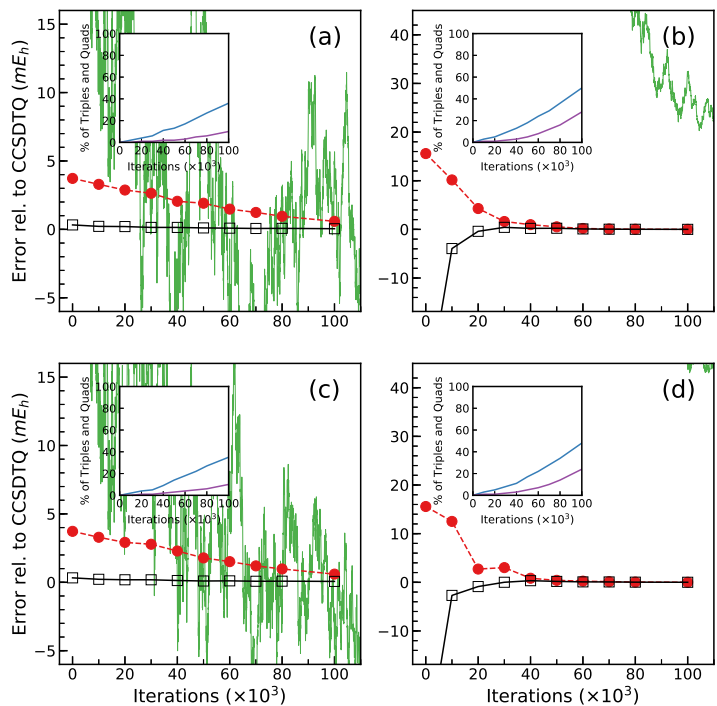}
\vspace*{-0.5em}
\caption{
Convergence of the CC($P$) (red filled circles and dashed lines) and
CC($P$;$Q$)$_{\rm EN}$ (black open squares and solid lines) energies toward
CCSDTQ for the water molecule, as described by the cc-pVDZ basis set.
The relevant $i$-CIQMC runs (all using $\delta\tau = 0.0001$ a.u.)
are depicted by the green lines representing the corresponding projected energies.
Panels (a) and (b) correspond to the calculations
in which the $P$ spaces employed in the CC($P$) steps consisted of all
singles and doubles and subsets of triples and quadruples identified during the $i$-FCIQMC
propagations; the $Q$ spaces needed to
define the corresponding $\delta(P;Q)$ corrections consisted of the triples that
were not captured by $i$-FCIQMC.
Panels (c) and (d) correspond to the calculations
in which the $P$ spaces employed in the CC($P$) steps consisted of all
singles and doubles and subsets of triples and quadruples identified during the $i$-CISDTQ-MC
propagations; in this case, the $Q$ spaces needed to
define the corresponding $\delta(P;Q)$ corrections consisted of the triples that
were not captured by $i$-CISDTQ-MC.
Panels (a) and (c) correspond to the equilibrium geometry. Panels (b) and (d)
correspond to the geometry in which both O--H bonds in water are simultaneously
stretched by a factor of 3 without changing the $\angle$(H--O--H) angle.
All reported energies are errors relative to CCSDTQ in millihartree.
The insets show the percentages of triples (blue line) and quadruples (purple line) captured
during the relevant $i$-CIQMC propagations.
}
\label{figure6}
\end{figure*}

Our last example, which illustrates the ability of the semi-stochastic CC($P$) and CC($P$;$Q$) approaches to
converge the CCSDTQ energetics, is the $C_{2v}$-symmetric cut of the ground-state potential energy surface of the
water molecule, in which both O--H bonds are simultaneously stretched without changing the $\angle$(H--O--H)
angle, resulting in large $T_{3}$ and $T_{4}$ contributions.
Following Ref. \onlinecite{olsen-h2o}, and consistent with our earlier deterministic CC($P$;$Q$) study,
\cite{nbjspp-molphys2017} where we also obtained the reference CCSDTQ energies, we used the spherical cc-pVDZ
basis set, correlated all electrons, and considered four stretches of the O--H bonds, including
$R_{\rm O\mbox{-}H} = 1.5 R_{e}$, $2 R_{e}$, $2.5 R_{e}$, and $3 R_{e}$, in addition to the equilibrium
geometry, $R_{\rm O\mbox{-}H} = R_{e}$. We used the same geometries, which the reader can find in
Ref. \onlinecite{olsen-h2o}, in the semi-stochastic CC($P$) and CC($P$;$Q$) calculations for ${\rm H_{2}O}$
carried out in this work, summarized in
Table \ref{table4} and Fig. \ref{figure6}. The authors of Ref. \onlinecite{olsen-h2o} obtained the
CCSDTQ energies too, but we rely on our own CCSDTQ data, published in Ref. \onlinecite{nbjspp-molphys2017}
and recalculated in this study, since Ref. \onlinecite{olsen-h2o} does not provide
the CCSDTQ results for $R_{\rm O\mbox{-}H} =2.5 R_{e}$ and $3 R_{e}$
and the CCSDTQ energies for $R_{\rm O\mbox{-}H} =1.5 R_{e}$ and $2 R_{e}$ reported
in Ref. \onlinecite{olsen-h2o} are in slight disagreement with the correctly converged values.

Up to twice the equilibrium O--H bond lengths, the CCSDT approach provides
an accurate description of the electronic energies of water, resulting in the 0.493, 1.423, and $-1.405$
millihartree signed errors relative to FCI at $R_{\rm O\mbox{-}H} = R_{e}$, $1.5 R_{e}$, and $2 R_{e}$,
respectively, when the cc-pVDZ basis set is employed, but when $R_{\rm O\mbox{-}H} > 2 R_{e}$, CCSDT
completely fails,\cite{nbjspp-molphys2017,olsen-h2o} and the CCSD(T), CCSD(2)$_{\rm T}$ or CR-CC(2,3)$_{\rm A}$
(in Table \ref{table4}, $\tau = 0$ CC($P$;$Q$)$_{\rm MP}$), CR-CC(2,3)$_{\rm D}$ (in Table \ref{table4},
$\tau = 0$ CC($P$;$Q$)$_{\rm EN}$), CCSDt, and CC(t;3) approximations to CCSDT, which were examined in Refs.
\onlinecite{nbjspp-molphys2017,crccl_jcp,olsen-h2o,ccsdpt2}, fail with it (CCSD(T) fails already at
$R_{\rm O\mbox{-}H} = 2 R_{e}$). In particular, the difference between the
CCSDT and FCI energies obtained with the cc-pVDZ basis set at $R_{\rm O\mbox{-}H} = 2.5 R_{e}$ is $-24.752$
millihartree. At $R_{\rm O\mbox{-}H} = 3 R_{e}$, the situation becomes even more dramatic, with the
CCSDT/cc-pVDZ energy falling 40.126 millihartree below its FCI counterpart.\cite{nbjspp-molphys2017,olsen-h2o}
One needs to incorporate $T_{4}$ clusters to reduce these massive errors in the $R_{\rm O\mbox{-}H} > 2 R_{e}$
region, and in order to do it in a reliable manner one has to use full CCSDTQ or one of the robust approximations
to it, such as the CCSDtq, CC(t,q;3), and CC(t,q;3,4) methods tested in Ref. \onlinecite{nbjspp-molphys2017}.
The conventional $T_{3}$ plus $T_{4}$ corrections to CCSD, such as CCSD(TQ$_{\rm f})$,\cite{ccsdtq-f} or their
CCSD(2)$_{\rm TQ}$\cite{eomccpt,ccsdpt2} and CR-CC(2,4) \cite{crccl_jcp,crccl_cpl,ptcp2007} counterparts examined
in Refs. \onlinecite{nbjspp-molphys2017,ccsdpt2} do not suffice. The CCSDT(2)$_{\rm Q}$ quadruples correction
to CCSDT\cite{eomccpt} is not robust enough either.\cite{ccsdpt2}

When the cc-pVDZ basis set is employed,
the differences between the CCSDTQ and FCI energies at $R_{\rm O\mbox{-}H} = R_{e}$,
$1.5 R_{e}$, $2 R_{e}$, $2.5 R_{e}$, and $3 R_{e}$ are 0.019, 0.121, 0.030, $-2.361$, and $-4.733$
millihartree, respectively,\cite{nbjspp-molphys2017}
which is a huge improvement over CCSDT. One might argue the need for
the inclusion of $T_{n}$ clusters with $n > 4$ at $R_{\rm O\mbox{-}H} = 2.5 R_{e}$ and $3 R_{e}$,
or try to obtain further improvements in describing the $R_{\rm O\mbox{-}H} > 2 R_{e}$ region
by replacing the RHF reference determinants used throughout this work by their unrestricted
counterparts, but studies of this kind are outside the scope of this article. The goal of the calculations
for the water molecule discussed in this subsection is to explore the potential offered by
the semi-stochastic CC($P$) and CC($P$;$Q$) approaches, especially the CC($P$;$Q$) corrections
to the CC($P$) energies calculated with the help of the FCIQMC and CISDTQ-MC propagations,
in converging the CCSDTQ energetics obtained with the spin- and symmetry-adapted RHF references.

As
demonstrated
in Table \ref{table4} and Fig. \ref{figure6},
which show the convergence of the CC($P$) and CC($P$;$Q$) energies toward their CCSDTQ parents,
and Table S.4 of the supplementary material, which reports the total numbers of walkers characterizing
the underlying CIQMC runs as percentages of the walker populations at $\tau = 10.0$ a.u., where our
CIQMC propagations were terminated,
the semi-stochastic CC($P$;$Q$)
calculations using FCIQMC and CISDTQ-MC are extremely efficient in capturing the combined effects
of $T_{3}$ and $T_{4}$ correlations.
This remains true
even in the most challenging $R_{\rm O\mbox{-}H} > 2 R_{e}$ region,
where the $T_{4}$ contributions, which have to overcome the massive failures of the CCSDT approach,
are very large and difficult to balance with their $T_{3}$ counterparts.
The FCIQMC- and CISDTQ-MC-driven CC($P$;$Q$) computations
accurately reproduce the parent CCSDTQ energetics already in the early stages of the
underlying CIQMC propagations,
when the stochastically determined $P$ spaces contain small fractions of triples and even smaller
fractions of quadruples and when the total numbers of walkers used in the CIQMC runs
are much smaller than those required to converge these runs. The FCIQMC- and CISDTQ-MC-based
CC($P$;$Q$) approaches greatly accelerate convergence of the corresponding
CC($P$) calculations, in spite of the fact
that in our current implementation of the semi-stochastic CC($P$;$Q$) routines aimed at CCSDTQ
the noniterative correction $\delta(P;Q)$ corrects the energy obtained by solving the CC($P$) equations
in the space of all singles and doubles and subsets of triples and quadruples captured by FCIQMC or CISDTQ-MC
for the triples outside the stochastically determined $P$ space, but not for the quadruples
missed by CIQMC.

Similarly to the previously discussed CC($P$;$Q$) calculations aimed at CCSDT, the
CC($P$;$Q$) approach targeting CCSDTQ that adopts the CC($P$;$Q$)$_{\rm EN}$ correction is generally most
effective, although the results of the CC($P$;$Q$)$_{\rm MP}$ calculations, in which the Epstein--Nesbet
denominator $D_{K}(P)$ in Eq. (\ref{eq:ell}) is replaced by its M{\o}ller--Plesset form, are as accurate
as their CC($P$;$Q$)$_{\rm EN}$ counterparts in the quasi-degenerate $R_{\rm O\mbox{-}H} > 2 R_{e}$ region.
Indeed, when we look at the
results in Table \ref{table4} corresponding to $R_{\rm O\mbox{-}H} = 2.5 R_{e}$ and $3 R_{e}$, where
the $T_{4}$ effects, estimated by forming the differences of the CCSDTQ and CCSDT energies, exceed
22 and 35 millihartree, respectively,\cite{nbjspp-molphys2017} and where the differences between the
CCSDT and CCSD energies, which measure the magnitude of $T_{3}$ contributions, are about $-45$
and $-51$ millihartree, respectively,\cite{nbjspp-molphys2017,olsen-h2o}
the FCIQMC- and CISDTQ-MC-based CC($P$;$Q$)$_{\rm EN}$ computations reduce the large $-20.739$
($R_{\rm O\mbox{-}H} = 2.5 R_{e}$) and $-35.823$ ($R_{\rm O\mbox{-}H} = 3 R_{e}$) millihartree
errors relative to CCSDTQ obtained in the initial CR-CC(2,3)$_{\rm D}$ ($\tau = 0$ CC($P$;$Q$)$_{\rm EN}$)
calculations to fractions of a millihartree after only 20,000 $\delta \tau = 0.0001$ a.u.
MC iterations, i.e., after the FCIQMC and CISDTQ-MC propagations capture as little as 5--6 \% of the
triples and 1 \% of the quadruples in the corresponding $P$ spaces. The FCIQMC- and CISDTQ-MC-driven
CC($P$;$Q$)$_{\rm MP}$ calculations using the same QMC propagation time $\tau$ are similarly effective
though. They reduce the large $-13.469$ and $-28.302$ millihartree errors relative to CCSDTQ resulting
from the initial CCSD(2)$_{\rm T}$ or CR-CC(2,3)$_{\rm A}$ ($\tau = 0$ CC($P$;$Q$)$_{\rm MP}$) computations
to a submillihartree level too. 

The situation changes in the $R_{\rm O\mbox{-}H} = R_{e} - 2 R_{e}$ region,
where the $T_{4}$ effects are much smaller than those originating from the $T_{3}$ clusters.
In this case, the convergence of the energies obtained in the semi-stochastic
CC($P$;$Q$)$_{\rm MP}$ calculations toward CCSDTQ is slower than that obtained with the
CC($P$;$Q$)$_{\rm EN}$ approach, i.e., our earlier conclusion, drawn from the calculations discussed
in Sections \ref{sec3.1} and \ref{sec3.2} and Ref. \onlinecite{stochastic-ccpq-prl-2017}, that the
use of the CC($P$;$Q$)$_{\rm EN}$ corrections to the semi-stochastic CC($P$) energies is generally
most effective still stands. This becomes particularly clear when we compare the results of the
FCIQMC- and CISDTQ-MC-driven CC($P$;$Q$)$_{\rm MP}$ and
CC($P$;$Q$)$_{\rm EN}$ calculations at $R_{\rm O\mbox{-}H} = R_{e}$ and $1.5R_{e}$. For example, it takes
only 40,000 $\delta \tau = 0.0001$ a.u. MC time steps, or about 10 \% of the triples and 2 \% of the
quadruples captured in the $P$ space, for the CC($P$;$Q$)$_{\rm EN}$ approach to reach a 0.1 millihartree
accuracy level relative to CCSDTQ at $R_{\rm O\mbox{-}H} = R_{e}$. The CC($P$;$Q$)$_{\rm MP}$ calculations
reach the same accuracy level after 100,000 MC time steps that capture about 35 \% of the triples and 10 \% of the
quadruples. When the $R_{\rm O\mbox{-}H} = 1.5 R_{e}$ geometry is considered, the CC($P$;$Q$)$_{\rm EN}$
calculations reach a 0.1 millihartree accuracy level relative to CCSDTQ after 60,000--70,000 MC iterations
that capture about 30 \% of the triples and 6--9 \% of the quadruples, i.e., in the relatively early stages
of the FCIQMC and CISDTQ-MC propagations. The CC($P$;$Q$)$_{\rm MP}$ calculations reach a similar accuracy
level 20,000--30,000 MC iterations later, after capturing about 40 \% of the triples and more than 10 \% of the
quadruples. It is certainly reassuring that the CC($P$;$Q$)$_{\rm EN}$ calculations using FCIQMC and
CISDTQ-MC to identify the leading triply and quadruply excited determinants for the inclusion in the
underlying $P$ spaces are capable of reproducing the CCSDTQ energies of the
water molecule over a wide range of geometries along the $C_{2v}$-symmetric cut of the ground-state
potential energy surface considered in Table \ref{table4} and Fig. \ref{figure6} to within $\sim$0.1 millihartree
out of the early stages of the CIQMC propagations, after capturing about 10 \% ($R_{\rm O\mbox{-}H} = R_{e}$)
or 30 \% ($R_{\rm O\mbox{-}H} > R_{e}$) of the triples and 2 \% ($R_{\rm O\mbox{-}H} = R_{e}$) or
about 10 \% ($R_{\rm O\mbox{-}H} > R_{e}$) of the quadruples. Having said this, it is interesting
to observe that both types of the CC($P$;$Q$) corrections tested in this study, abbreviated as
CC($P$;$Q$)$_{\rm MP}$ and CC($P$;$Q$)$_{\rm EN}$,
perform equally well when $R_{\rm O\mbox{-}H} > 2 R_{e}$, i.e., when the $T_{3}$ and $T_{4}$ effects
are both very large. We observed a similar behavior in Ref. \onlinecite{nbjspp-molphys2017}, when examining
the relative performance of the CC($P$;$Q$)-based CC(t,q;3)$_{\rm A}$ and CC(t,q;3)$_{\rm D}$ corrections
to CCSDtq using the double dissociation of water as one of the examples. This should not be surprising, since
the CC(t,q;3)$_{\rm A}$ and CC(t,q;3)$_{\rm D}$ methods investigated in Ref. \onlinecite{nbjspp-molphys2017}
can be regarded as the deterministic counterparts of the semi-stochastic CC($P$;$Q$)$_{\rm MP}$ and
CC($P$;$Q$)$_{\rm EN}$ approaches targeting the CCSDTQ energetics implemented in this work.

As in the case of the CC($P$;$Q$) calculations targeting CCSDT, discussed in Sections \ref{sec3.1}
and \ref{sec3.2}, the observed fast convergence of the semi-stochastic CC($P$;$Q$) calculations aimed
at recovering the CCSDTQ energetics does not seem to be affected by the type of the CIQMC approach
used to identify the leading triply and quadruply excited determinants.
This should facilitate future applications of the semi-stochastic CC($P$;$Q$)
methodology, including cases of stronger electronic quasi-degeneracies characterized by large
$T_{3}$ and $T_{4}$ contributions, helping us to converge the CCSDTQ-level energetics at the
small fraction of the deterministic CCSDTQ effort by taking advantage of the least expensive forms
CIQMC capable of capturing triples and quadruples, represented in this study by CISDTQ-MC.

\section{Conclusions}
\label{sec4}

We have recently started exploring a novel way of obtaining accurate electronic energetics
equivalent to high-level CC calculations, at the small fraction of the computational
effort and preserving the black-box character of conventional single-reference
computations, by merging the deterministic CC($P$;$Q$) formalism, originally proposed
in Refs. \onlinecite{jspp-chemphys2012,jspp-jcp2012}, along with the underlying
CC($P$)/EOMCC($P$) framework, with the stochastic CIQMC
\cite{Booth2009,Cleland2010,fciqmc-uga-2019,ghanem_alavi_fciqmc_jcp_2019} and CCMC
\cite{Thom2010,Franklin2016,Spencer2016,Scott2017} approaches.
\cite{stochastic-ccpq-prl-2017,eomccp-jcp-2019,stochastic-ccpq-molphys-2020}
When combined with the FCIQMC and CCSDT-MC wave function sampling, used to identify the
leading triply excited determinants or cluster/excitation amplitudes, and correcting the
CC($P$)\cite{stochastic-ccpq-prl-2017} and EOMCC($P$)\cite{eomccp-jcp-2019} energies for the remaining
triples not captured by FCIQMC or CCSDT-MC, the resulting semi-stochastic CC($P$;$Q$) methodology
\cite{stochastic-ccpq-prl-2017} and its excited-state extension\cite{stochastic-ccpq-molphys-2020}
turned out to be very promising, allowing us to converge the CCSDT and EOMCCSDT
energetics out of the early stages of the underlying QMC propagations.

This study can be regarded as the next key step in the development and exploration of
the semi-stochastic CC($P$;$Q$) approaches, in which we have extended our initial work,
\cite{stochastic-ccpq-prl-2017} focusing on recovering the CCSDT energetics and
relying on FCIQMC and CCSDT-MC, to more efficient ways of identifying the leading
higher--than--doubly excited determinants for the inclusion in the
underlying $P$ spaces. We have accomplished this goal by replacing FCIQMC by its less
expensive CISDT-MC and CISDTQ-MC counterparts. We have also developed and tested the
initial variant of the semi-stochastic CC($P$;$Q$) method aimed at converging
the CCSDTQ energetics, in which the results of CC($P$) calculations in the
subspaces spanned by singles, doubles, and subsets of triples and quadruples
identified by FCIQMC or CISDTQ-MC are corrected for the remaining triples outside
the stochastically determined $P$ spaces. By comparing the FCIQMC-driven CC($P$;$Q$) calculations
targeting CCSDT, carried out in this work, in which the noniterative corrections $\delta(P;Q)$
to the CC($P$) energies have been treated fully, as required by Eqs. (\ref{mmcc-gen-delta}),
(\ref{eq:ell}), and (\ref{eq:denom}), with the analogous computations
reported in Ref. \onlinecite{stochastic-ccpq-prl-2017}, where the same corrections
were treated in a somewhat simplified manner by neglecting the three-body component of
the deexcitation operator ${\Lambda}^{(P)}$ used to construct amplitudes $\ell_{K}(P)$
entering Eq. (\ref{mmcc-gen-delta}), we have examined the significance of the
full {\it vs} approximate treatment of these corrections for the accuracy of the resulting
CC($P$;$Q$) energies. Other important issues, such as the benefits of using the
Epstein--Nesbet form of the denominators $D_{K}(P)$ that enter the definition
of corrections $\delta(P;Q)$, resulting in the CC($P$;$Q$)$_{\rm EN}$ variant of
CC($P$;$Q$), as compared to their M{\o}ller--Plesset counterparts defining the
CC($P$;$Q$)$_{\rm MP}$ corrections, have been investigated as well.

The ability of the semi-stochastic CC($P$;$Q$) approaches to converge the CCSDT and
CCSDTQ energies, based on the truncated CISDT-MC and CISDTQ-MC propagations, and their
FCIQMC counterparts in which the noniterative corrections $\delta(P;Q)$ have been
treated fully, has been illustrated using a few molecular examples, for which
the deterministic CCSDT and CCSDTQ calculations that provide the reference data
are feasible and which require a high-level CC treatment to obtain a reliable description.
Thus, we have reported the results of the semi-stochastic CC($P$;$Q$) calculations
using CISDT-MC, CISDTQ-MC, and FCIQMC aimed at converging the CCSDT energetics
for the F--F bond breaking in ${\rm F}_{2}$ and the automerization of cyclobutadiene,
which require an accurate treatment of $T_{3}$ clusters accounting for
the relaxation of $T_{1}$ and $T_{2}$ amplitudes in the presence of large $T_{3}$
contributions, and the CISDTQ-MC- and FCIQMC-driven CC($P$;$Q$) computations
for the $C_{2v}$-symmetric stretching of the O--H bonds in the water molecule
targeting CCSDTQ, where the $T_{3}$ and $T_{4}$ clusters become
large and difficult to balance.

The numerical results reported in this article clearly show that the semi-stochastic CC($P$;$Q$)
calculations are capable of accurately reproducing the parent CCSDT and CCSDTQ energetics, even when
electronic quasi-degeneracies and higher--than--two-body components of the cluster operator become large, out
of the early stages of the corresponding CIQMC propagations, accelerating convergence of the underlying
CC($P$) computations at the same time. The convergence of the CC($P$;$Q$) energies toward their
CCSDT and CCSDTQ parents does not seem to be affected by the type of the CIQMC approach used to identify the
leading triply or triply and quadruply excited determinants. In the case of the CC($P$;$Q$) calculations
targeting the CCSDT energetics, one can use FCIQMC or one of its less expensive truncated forms, such as
CISDTQ-MC, or even the crude CISDT-MC approach, with virtually no impact on the systematic convergence pattern
toward CCSDT as the propagation time $\tau$ approaches $\infty$.
Similarly, one can replace FCIQMC by CISDTQ-MC without any significant effect on the convergence
of the semi-stochastic CC($P$;$Q$) calculations toward CCSDTQ. Our calculations also suggest that a complete
treatment of the CC($P$;$Q$) corrections $\delta(P;Q)$, as defined by Eqs. (\ref{mmcc-gen-delta}),
(\ref{eq:ell}), and (\ref{eq:denom}), including the use of the CC($P$;$Q$)$_{\rm EN}$ approach, as opposed
to its more approximate CC($P$;$Q$)$_{\rm MP}$ version, is more important than the actual
type of the CIQMC approach used to determine the relevant $P$ spaces,
especially when one is interested in accelerating convergence
of the semi-stochastic CC($P$;$Q$) calculations in the early stages of the QMC propagations.
We have demonstrated that independent of the type of the CIQMC approach used to identify
the leading triply or triply and quadruply excited determinants for the inclusion in the
relevant $P$ spaces and independent of the magnitude of $T_{3}$ and $T_{4}$ effects,
the semi-stochastic CC($P$;$Q$) calculations allow us to reach submillihartree
accuracy levels relative to the parent CCSDT and CCSDTQ energetics with small fractions of
higher--than--doubly excited determinants captured in the early stages of the corresponding CIQMC runs
and with small walker populations that are far less than the total numbers of walkers required to
converge these runs.

By relaxing $T_{1}$ and $T_{2}$ clusters in the presence of their $T_{3}$ or
$T_{3}$ and $T_{4}$ counterparts defined using the excitation lists provided
by full or truncated CIQMC, the semi-stochastic CC($P$;$Q$) computations are
capable of considerably improving accuracy of the more established noniterative
corrections to CCSD without making the calculations a lot more expensive. In this sense,
the semi-stochastic CC($P$;$Q$) methodology using CIQMC is very
similar to the deterministic CC(t;3), CC(t,q;3), and CC(t,q;3,4) hierarchy developed and tested in Refs.
\onlinecite{jspp-chemphys2012,jspp-jcp2012,jspp-jctc2012,nbjspp-molphys2017,ccpq-be2-jpca-2018,ccpq-mg2-mp-2019},
which uses the CC($P$;$Q$) corrections to correct the results of the active-space CCSDt or CCSDtq calculations for
the remaining $T_{3}$ or $T_{3}$ and $T_{4}$ correlations that were not captured via active orbitals.
There is, however, one major advantage of the semi-stochastic CC($P$;$Q$) framework over the
CC(t;3), CC(t,q;3), and CC(t,q;3,4) approaches, namely, the use of FCIQMC or truncated CIQMC propagations, which
can efficiently identify the leading higher--than--doubly-excited determinants for the inclusion in the
relevant $P$ spaces, combined with the $\delta(P;Q)$ corrections to capture the remaining correlations of
interest, offers an automated way of performing accurate CC($P$;$Q$) computations without any reference to
the user- and system-dependent active orbitals. The analogies between the active-space CCSDt
(for excited states, EOMCCSDt\cite{eomkkpp,eomccsdt1,eomccsdt2})
and semi-stochastic CC($P$)/EOMCC($P$) approaches, on which the deterministic CC(t;3) (in the case
of CCSDt/EOMCCSDt) and CIQMC-driven (in the case of semi-stochastic CC($P$)/EOMCC($P$)) CC($P$;$Q$)
approaches are based, have been investigated in Ref. \onlinecite{stochastic-ccpq-molphys-2020}.

The findings presented in this article are encouraging from the point of view of future
applications of the semi-stochastic CC($P$;$Q$) methodology using CIQMC, including
challenging cases of stronger electronic quasi-degeneracies characterized by large $T_{3}$ or
$T_{3}$ and $T_{4}$ contributions that other approximations to CCSDT or CCSDTQ may struggle with,
but the story is not over yet. We certainly need to improve the efficiency of our CC($P$;$Q$)
codes, especially the underlying CC($P$) routines, to obtain full benefits offered by the
semi-stochastic CC($P$;$Q$) approaches, discussed in Section \ref{sec2.2}. This is especially
true in the case of our current CC($P$;$Q$) codes aimed at converging the CCSDTQ energetics,
which have a largely pilot character. In this case, we also need to examine if one can further
improve the convergence of the FCIQMC- or CISDTQ-MC-driven CC($P$;$Q$) calculations aimed at
CCSDTQ by correcting the underlying CC($P$) energies for both the missing triples and quadruples
not captured by CIQMC at a given time $\tau$, not just for the missing triples, as has been done
in this work. It would also be useful to examine if one can extend the semi-stochastic CC($P$)
and CC($P$;$Q$) approaches to the higher CC theory levels, beyond CCSDTQ examined in this work
and beyond EOMCCSDT explored in Refs. \onlinecite{eomccp-jcp-2019,stochastic-ccpq-molphys-2020},
and investigate if our observations regarding the utility of the truncated CIQMC methods, such
as CISDT-MC and CISDTQ-MC, remain true in the excited-state and open-shell CC($P$;$Q$) calculations.
In this study, we have adopted the original form of the $i$-CIQMC algorithm proposed in Ref.
\onlinecite{Cleland2010}, but it would be interesting to examine if one could obtain additional benefits
by interfacing our semi-stochastic CC($P$;$Q$) methods with the improved ways of converging CIQMC,
such as the adaptive-shift approach developed  Refs. \onlinecite{ghanem_alavi_fciqmc_jcp_2019,ghanem_alavi_fciqmc_2020}.
All of the above ideas are presently pursued in our group, and the results will be reported as
soon as they become available. Last, but not least, we have recently interfaced our CC($P$) and CC($P$;$Q$) routines
with some of the modern versions of the selected CI approaches, which date back to the late 1960s and
early 1970s \cite{sci_1,sci_2,sci_3,sci_4} and which have recently regained significant attention.
\cite{adaptive_ci_1,adaptive_ci_2,asci_1,asci_2,ici_1,ici_2,shci_1,shci_2,shci_3,cipsi_1,cipsi_2}
Our initial numerical results, which we hope to report in a separate publication,\cite{cipsi-ccpq-2021}
indicate that selected CI methods can be as effective in generating meaningful $P$ spaces for the CC($P$)
calculations, which precede the determination of the $\delta(P;Q)$ moment corrections, as the stochastic
CIQMC propagations advocated in this and our earlier\cite{stochastic-ccpq-prl-2017,eomccp-jcp-2019,stochastic-ccpq-molphys-2020}
studies.

\section*{Supplementary Material}

See the supplementary material for the information about the total numbers of walkers characterizing the
FCIQMC, CISDT-MC, and CISDTQ-MC propagations for the F--F bond breaking in ${\rm F}_{2}$ and
the automerization of cyclobutadiene and the FCIQMC and CISDTQ-MC propagations for the double
dissociation of the water molecule carried out in the present study.

\begin{acknowledgments}
This work has been supported by the Chemical Sciences, Geosciences
and Biosciences Division, Office of Basic Energy Sciences, Office
of Science, U.S. Department of Energy (Grant No. DE-FG02-01ER15228 to P.P),
the National Science Foundation (Grant No. CHE-1763371 to P.P.), and Phase
I and II Software Fellowships awarded to J.E.D. by the Molecular Sciences Software Institute
funded by the National Science Foundation grant ACI-1547580.
P.P. thanks Professors Ali Alavi, George H. Booth, and Alex J. W. Thom
for useful discussions.
\end{acknowledgments}

\section*{Data Availability}

The data that support the findings of this study are available within the article
and its supplementary material.

\vspace*{1em}

%

\pagebreak


\onecolumngrid

\squeezetable
\begin{table*}[h!]
\caption{
\label{table1}
Convergence of the CC($P$), CC($P$;$Q$)$_{\rm MP}$, and CC($P$;$Q$)$_{\rm EN}$
energies toward CCSDT, where the $P$ spaces consisted of all singles
and doubles and subsets of triples identified during the $i$-FCIQMC, $i$-CISDTQ-MC, or $i$-CISDT-MC
propagations with $\delta\tau = 0.0001$ a.u. and where the corresponding $Q$ spaces
consisted of the triples not captured by
the corresponding QMC simulations, for the ${\rm F}_{2}$/cc-pVDZ molecule
in which the F--F distance $R$ was set at $R_{e}$, $1.5 R_{e}$, $2 R_{e}$, and $5 R_{e}$,
with $R_{e} = 2.66816$ bohr representing the equilibrium geometry.
The $i$-FCIQMC, $i$-CISDTQ-MC, and $i$-CISDT-MC calculations preceding the CC($P$) and
CC($P$;$Q$) steps were initiated by placing 100 walkers on the RHF determinant
and the $n_{a}$ parameter of the initiator algorithm was set at 3.
In all post-RHF calculations, the lowest two core orbitals were
kept frozen and the Cartesian components of $d$ orbitals were employed throughout.
}
\begin{ruledtabular}
\begin{tabular}{lccccddddddddd}
  &  & \multicolumn{3}{c}{\textrm{\% of triples}} & \multicolumn{3}{c}{\textrm{CC($P$)\footnotemark[1]}} & \multicolumn{3}{c}{\textrm{CC($P$;$Q$)$_{\rm MP}$\footnotemark[1]}} & \multicolumn{3}{c}{\textrm{CC($P$;$Q$)$_{\rm EN}$\footnotemark[1]}} \\ \cline{3-5} \cline{6-8} \cline{9-11} \cline{12-14}
$R/R_{e}$ & MC iterations & \multicolumn{1}{c}{FCI\footnotemark[2]} & \multicolumn{1}{c}{CIQ\footnotemark[3]} & \multicolumn{1}{c}{CIT\footnotemark[4]} & \multicolumn{1}{c}{FCI\footnotemark[2]} & \multicolumn{1}{c}{CIQ\footnotemark[3]} & \multicolumn{1}{c}{CIT\footnotemark[4]}& \multicolumn{1}{c}{FCI\footnotemark[2]} & \multicolumn{1}{c}{CIQ\footnotemark[3]} & \multicolumn{1}{c}{CIT\footnotemark[4]} & \multicolumn{1}{c}{FCI\footnotemark[2]} & \multicolumn{1}{c}{CIQ\footnotemark[3]} & \multicolumn{1}{c}{CIT\footnotemark[4]}\\
  \hline
  1.0 & 0 & \multicolumn{3}{c}{0} & \multicolumn{3}{d}{9.485\footnotemark[5]} & \multicolumn{3}{d}{1.398\footnotemark[6]} & \multicolumn{3}{d}{-0.240\footnotemark[7]} \\
&10000	&3	& 3 & 4 &5.692 & 5.692&5.229&0.760& 0.760& 0.688 &-0.151&-0.151 &-0.152 \\
&20000	&9	& 8 & 8 &3.548 & 3.804&3.962&0.444& 0.473& 0.472 &-0.107&-0.093 &-0.140 \\
&30000	&15	& 16& 14&2.290 & 2.498&2.769&0.284& 0.301& 0.334 &-0.059&-0.046 &-0.067 \\
&40000	&25	& 26& 22&1.791 & 1.523&1.765&0.212& 0.184& 0.210 &-0.037&-0.030 &-0.034 \\
&50000	&37	& 38& 34&0.933 & 0.940&1.151&0.113& 0.115& 0.137 &-0.014&-0.013 &-0.021 \\
&60000	&51	& 52& 46&0.536 & 0.498&0.698&0.064& 0.058& 0.083 &-0.008&-0.008 &-0.010 \\
&70000	&63	& 64& 58&0.383 & 0.308&0.410&0.044& 0.036& 0.047 &-0.006&-0.004 &-0.007 \\
&80000	&73	& 74& 68&0.177 & 0.164&0.224&0.020& 0.018& 0.025 &-0.003&-0.002 &-0.003 \\
&100000	&89	& 89& 85&0.044 & 0.050&0.073&0.005& 0.006& 0.008 & 0.000&-0.001 &-0.001 \\
&120000	&97	& 97& 94&0.013 & 0.010&0.024&0.001& 0.001& 0.003 & 0.000& 0.000 & 0.000 \\
    & $\infty$ & \multicolumn{3}{c}{100} & \multicolumn{3}{d}{-199.102796\footnotemark[8]} & \multicolumn{3}{c}{---} & \multicolumn{3}{c}{---} \\
  &&&&&&&&\\
  1.5 & 0 & \multicolumn{3}{c}{0} & \multicolumn{3}{d}{32.424\footnotemark[5]} & \multicolumn{3}{d}{5.984\footnotemark[6]} & \multicolumn{3}{d}{1.735\footnotemark[7]} \\
&10000  &3	&3  & 3    &14.312&14.220  &  15.874&2.198 & 1.980& 2.115 & 0.351 & 0.321&  0.193  \\
&20000  &9	&8  & 7    &5.589	&3.572   &  5.564 &0.629 & 0.428& 0.657 &-0.003 &-0.000&  0.052  \\
&30000  &16	&18 & 14   &2.728	&2.391   &  2.206 &0.323 & 0.285& 0.262 &-0.002 & 0.020&  0.021  \\
&40000  &27	&30 & 24   &1.065	&0.706   &  1.387 &0.142 & 0.084& 0.171 & 0.020 & 0.009&  0.015  \\
&50000  &42	&45 & 35   &0.482	&0.459   &  0.687 &0.062 & 0.055& 0.087 & 0.009 & 0.006&  0.008  \\
&60000  &57	&60 & 49   &0.273	&0.219   &  0.336 &0.029 & 0.027& 0.041 & 0.001 & 0.000&  0.005  \\
&70000  &70	&72 & 61   &0.128	&0.106   &  0.231 &0.013 & 0.011& 0.028 & 0.000 & 0.000&  0.001  \\
&80000  &81	&82 & 72   &0.064	&0.048   &  0.102 &0.006 & 0.004& 0.010 & 0.000 &-0.001& -0.001  \\
&100000	&93	&94 & 88   &0.012	&0.009   &  0.026 &0.001 & 0.001& 0.003 & 0.000 & 0.000&  0.000  \\
&120000	&99	&100& 96   &0.001	&0.002   &  0.005 &0.000 & 0.000& 0.000 & 0.000 & 0.000&  0.000  \\
  & $\infty$ & \multicolumn{3}{c}{100} & \multicolumn{3}{d}{-199.065882\footnotemark[8]} & \multicolumn{3}{c}{---} & \multicolumn{3}{c}{---} \\
  &&&&&\\
  2.0 & 0 & \multicolumn{3}{c}{0} & \multicolumn{3}{d}{45.638\footnotemark[5]} & \multicolumn{3}{d}{6.357\footnotemark[6]} & \multicolumn{3}{d}{1.862\footnotemark[7]} \\
&10000	&4	&4  & 3  &12.199  &17.779  & 12.687&0.998&1.886 &1.181&-0.063 & 0.280&-0.008\\
&20000	&10	&9  & 9  &4.127   &2.529   & 3.672 &0.328&0.245 &0.310&-0.014 & 0.009&-0.025\\
&30000	&21	&19 & 17 &0.802   &1.172   & 1.393 &0.081&0.115 &0.128& 0.008 & 0.011& 0.004\\
&40000	&35	&32 & 28 &0.456   &0.499   & 0.627 &0.040&0.047 &0.058&-0.001 & 0.000& 0.000\\
&50000	&51	&48 & 41 &0.216   &0.215   & 0.305 &0.018&0.019 &0.027&-0.001 & 0.000&-0.001\\
&60000	&66	&64 & 56 &0.083   &0.112   & 0.160 &0.007&0.010 &0.014&-0.001 &-0.001&-0.001\\
&70000	&79	&75 & 68 &0.037   &0.048   & 0.074 &0.003&0.004 &0.006& 0.000 &-0.001&-0.001\\
&80000	&87	&85 & 78 &0.013   &0.019   & 0.034 &0.001&0.002 &0.003& 0.000 & 0.000& 0.000\\
&100000	&97	&95 & 91 &0.001   &0.002   & 0.007 &0.000&0.000 &0.001& 0.000 & 0.000& 0.000\\
&120000	&100&100& 98 &0.000   &0.000   & 0.000 &0.000&0.000 &0.000& 0.000 & 0.000& 0.000\\
& $\infty$ & \multicolumn{3}{c}{100} & \multicolumn{3}{d}{-199.058201\footnotemark[8]} & \multicolumn{3}{c}{---} & \multicolumn{3}{c}{---} \\
  &&&&&\\
  5.0 & 0 & \multicolumn{3}{c}{0} & \multicolumn{3}{d}{49.816\footnotemark[5]} & \multicolumn{3}{d}{3.895\footnotemark[6]} & \multicolumn{3}{d}{1.613\footnotemark[7]} \\
&10000	&3	&3 & 3  &10.887	& 13.326   & 9.776 &0.455&0.672&0.642 &-0.005& 0.059& 0.202 \\
&20000	&8	&8 & 8  &1.968	& 2.535    & 1.315 &0.152&0.165&0.102 & 0.040& 0.026& 0.012 \\
&30000	&17	&15& 15 &0.529	& 0.752    & 1.042 &0.041&0.056&0.081 & 0.001& 0.006& 0.015 \\
&40000	&27	&26& 26 &0.295	& 0.351    & 0.346 &0.022&0.024&0.025 & 0.001&-0.001&-0.001 \\
&50000	&38	&37& 36 &0.116	& 0.147    & 0.166 &0.008&0.011&0.011 &-0.001& 0.000&-0.001 \\
&60000	&47	&46& 44 &0.047	& 0.059    & 0.070 &0.003&0.004&0.005	&-0.001& 0.000&-0.001 \\
&70000	&54	&52& 50 &0.016	& 0.020    & 0.030 &0.001&0.001&0.002	& 0.000& 0.000& 0.000 \\
&80000	&60	&59& 55 &0.006	& 0.006    & 0.014 &0.000&0.000&0.001	& 0.000& 0.000& 0.000 \\
&100000	&74	&73& 66 &0.000	& 0.000    & 0.001 &0.000&0.000&0.000	& 0.000& 0.000& 0.000 \\
&120000	&89	&87& 78 &0.000	& 0.000    & 0.000 &0.000&0.000&0.000	& 0.000& 0.000& 0.000 \\
& $\infty$ & \multicolumn{3}{c}{100} & \multicolumn{3}{d}{-199.058586\footnotemark[8]} & \multicolumn{3}{c}{---} & \multicolumn{3}{c}{---} \\
\end{tabular}
\end{ruledtabular}
\footnotetext[1]{
\setlength{\baselineskip}{1em}
Unless otherwise stated, all energies are reported as errors relative to CCSDT in millihartree.
}
\footnotetext[2]{
\setlength{\baselineskip}{1em}
FCI stands for \textit{i}-FCIQMC.
}
\footnotetext[3]{
\setlength{\baselineskip}{1em}
CIQ stands for \textit{i}-CISDTQ-MC.
}
\footnotetext[4]{
\setlength{\baselineskip}{1em}
CIT stands for \textit{i}-CISDT-MC.
}
\footnotetext[5]{
\setlength{\baselineskip}{1em}
Equivalent to CCSD.
}
\footnotetext[6]{
\setlength{\baselineskip}{1em}
Equivalent to the CCSD energy corrected for the effects of $T_{3}$ clusters using
the CCSD(2)$_{\rm T}$ approach of Ref. \onlinecite{ccsdpt2}, which is equivalent to the approximate
form of the completely renormalized CR-CC(2,3) approach of Refs. \onlinecite{crccl_jcp,crccl_cpl},
abbreviated sometimes as CR-CC(2,3),A or CR-CC(2,3)$_{\rm A}$.
\cite{jspp-jctc2012,nbjspp-molphys2017,crccl_jpc,ptcp2007,crccl_ijqc2}
}
\footnotetext[7]{
\setlength{\baselineskip}{1em}
Equivalent to the CCSD energy corrected for the effects of $T_{3}$ clusters using the
most complete variant of the completely renormalized CR-CC(2,3) approach of Refs. \onlinecite{crccl_jcp,crccl_cpl},
abbreviated sometimes as CR-CC(2,3),D or CR-CC(2,3)$_{\rm D}$.
\cite{jspp-jctc2012,nbjspp-molphys2017,crccl_jpc,ptcp2007,crccl_ijqc2}
}
\footnotetext[8]{
\setlength{\baselineskip}{1em}
Total CCSDT energy in hartree.
}
\end{table*}

\begin{table*}[h!]
\caption{
\label{table2}
Convergence of the CC($P$), CC($P$;$Q$)$_{\rm MP}$, and CC($P$;$Q$)$_{\rm EN}$
energies toward CCSDT, where the $P$ spaces consisted of all singles
and doubles and subsets of triples identified during the $i$-FCIQMC, $i$-CISDTQ-MC, or $i$-CISDT-MC
propagations with $\delta\tau = 0.0001$ a.u. and where the corresponding $Q$ spaces
consisted of the triples not captured by
the corresponding QMC simulations, for the ${\rm F}_{2}$ molecule
in which the F--F distance $R$ was set at twice the equilibrium bond length,
using the cc-pVTZ and aug-cc-pVTZ basis sets, abbreviated as VTZ and AVTZ, respectively.
The $i$-FCIQMC, $i$-CISDTQ-MC, and $i$-CISDT-MC calculations preceding the CC($P$) and
CC($P$;$Q$) steps were initiated by placing 100 walkers on the RHF determinant
and the $n_{a}$ parameter of the initiator algorithm was set at 3.
In all post-RHF calculations, the lowest two core orbitals were
kept frozen and the spherical components of $d$ and $f$ orbitals were employed throughout.
}
\begin{ruledtabular}
\begin{tabular}{lccccddddddddd}
  &  & \multicolumn{3}{c}{\textrm{\% of triples}} & \multicolumn{3}{c}{\textrm{CC($P$)\footnotemark[1]}} & \multicolumn{3}{c}{\textrm{CC($P$;$Q$)$_{\rm MP}$\footnotemark[1]}} & \multicolumn{3}{c}{\textrm{CC($P$;$Q$)$_{\rm EN}$\footnotemark[1]}} \\ \cline{3-5} \cline{6-8} \cline{9-11} \cline{12-14}
  Basis set & MC iterations & \multicolumn{1}{c}{FCI\footnotemark[2]} & \multicolumn{1}{c}{CIQ\footnotemark[3]} & \multicolumn{1}{c}{CIT\footnotemark[4]} & \multicolumn{1}{c}{FCI\footnotemark[2]} & \multicolumn{1}{c}{CIQ\footnotemark[3]} & \multicolumn{1}{c}{CIT\footnotemark[4]}& \multicolumn{1}{c}{FCI\footnotemark[2]} & \multicolumn{1}{c}{CIQ\footnotemark[3]} & \multicolumn{1}{c}{CIT\footnotemark[4]} & \multicolumn{1}{c}{FCI\footnotemark[2]} & \multicolumn{1}{c}{CIQ\footnotemark[3]} & \multicolumn{1}{c}{CIT\footnotemark[4]}\\
  \hline
VTZ & 0 & \multicolumn{3}{c}{0} & \multicolumn{3}{d}{62.819\footnotemark[5]} & \multicolumn{3}{d}{9.211\footnotemark[6]} & \multicolumn{3}{d}{4.254\footnotemark[7]} \\
&10000  &1  &1  &1  &29.714 &31.973 &31.571 &2.738&3.104&2.636& 0.728& 0.896& 0.539\\
&20000  &2  &2  &2  &11.179 &14.687 &20.194 &0.824&1.097&1.487& 0.071& 0.151& 0.217\\
&30000  &6  &6  &4  &5.787  &6.031  &9.294  &0.400&0.425&0.617& 0.028& 0.030& 0.025\\
&40000  &14 &14 &10 &2.406  &2.574  &4.203  &0.160&0.171&0.284& 0.002& 0.001& 0.014\\
&50000  &27 &26 &19 &1.193  &1.237  &2.177  &0.076&0.078&0.138&-0.003&-0.002&-0.002\\
&60000  &42 &42 &30 &0.490  &0.489  &1.144  &0.029&0.029&0.071&-0.002&-0.002&-0.005\\
&70000  &59 &57 &44 &0.178  &0.171  &0.576  &0.011&0.010&0.037&-0.001&-0.001&-0.002\\
&80000  &72 &71 &56 &0.045  &0.054  &0.309  &0.003&0.003&0.020& 0.000& 0.000&-0.001\\
&100000 &90 &89 &78 &0.002  &0.003  &0.130  &0.000&0.000&0.009& 0.000& 0.000& 0.000\\
  & $\infty$ & \multicolumn{3}{c}{100} & \multicolumn{3}{d}{-199.238344\footnotemark[8]} & \multicolumn{3}{c}{---} & \multicolumn{3}{c}{---} \\
  &&&&&&&&\\
AVTZ & 0 & \multicolumn{3}{c}{0} & \multicolumn{3}{d}{65.036\footnotemark[5]} & \multicolumn{3}{d}{9.808\footnotemark[6]} & \multicolumn{3}{d}{5.595\footnotemark[7]} \\
&10000	&0	&0	&0	&36.316	&38.874	&42.801	&3.641&4.144&4.851& 1.594& 1.786& 2.304\\
&20000	&1	&1	&1	&17.190	&20.799	&26.557	&1.276&1.656&2.288& 0.382& 0.512& 0.791\\
&30000	&4	&4	&3	&8.065	&9.272	&13.279	&0.549&0.623&0.928& 0.138& 0.138& 0.246\\
&40000	&10	&10	&7	&4.408	&4.677	&7.477	&0.291&0.307&0.499& 0.057& 0.062& 0.106\\
&50000	&23	&22	&15	&2.208	&2.425	&3.951	&0.136&0.150&0.244& 0.016& 0.019& 0.038\\
&60000	&41	&39	&27	&1.021	&1.137	&2.052	&0.058&0.070&0.124& 0.002& 0.005& 0.013\\
&70000	&61	&58	&61	&0.385	&0.455	&0.385	&0.021&0.025&0.059& 0.000& 0.000& 0.001\\
&80000	&78	&76	&78	&0.125	&0.154	&0.125	&0.007&0.008&0.026& 0.000& 0.000& 0.000\\
&100000	&97	&96	&97	&0.007	&0.009	&0.007	&0.000&0.001&0.004& 0.000& 0.000& 0.000\\
& $\infty$ & \multicolumn{3}{c}{100} & \multicolumn{3}{d}{-199.253022\footnotemark[8]} & \multicolumn{3}{c}{---} & \multicolumn{3}{c}{---} \\
\end{tabular}
\end{ruledtabular}
\footnotetext[1]{
\setlength{\baselineskip}{1em}
Unless otherwise stated, all energies are reported as errors relative to CCSDT in millihartree.
}
\footnotetext[2]{
\setlength{\baselineskip}{1em}
FCI stands for \textit{i}-FCIQMC.
}
\footnotetext[3]{
\setlength{\baselineskip}{1em}
CIQ stands for \textit{i}-CISDTQ-MC.
}
\footnotetext[4]{
\setlength{\baselineskip}{1em}
CIT stands for \textit{i}-CISDT-MC.
}
\footnotetext[5]{
\setlength{\baselineskip}{1em}
Equivalent to CCSD.
}
\footnotetext[6]{
\setlength{\baselineskip}{1em}
Equivalent to the CCSD energy corrected for the effects of $T_{3}$ clusters using
the CCSD(2)$_{\rm T}$ approach of Ref. \onlinecite{ccsdpt2}, which is equivalent to the approximate
form of the completely renormalized CR-CC(2,3) approach of Refs. \onlinecite{crccl_jcp,crccl_cpl},
abbreviated sometimes as CR-CC(2,3),A or CR-CC(2,3)$_{\rm A}$.
\cite{jspp-jctc2012,nbjspp-molphys2017,crccl_jpc,ptcp2007,crccl_ijqc2}
}
\footnotetext[7]{
\setlength{\baselineskip}{1em}
Equivalent to the CCSD energy corrected for the effects of $T_{3}$ clusters using the
most complete variant of the completely renormalized CR-CC(2,3) approach of Refs. \onlinecite{crccl_jcp,crccl_cpl},
abbreviated sometimes as CR-CC(2,3),D or CR-CC(2,3)$_{\rm D}$.
\cite{jspp-jctc2012,nbjspp-molphys2017,crccl_jpc,ptcp2007,crccl_ijqc2}
}
\footnotetext[8]{
\setlength{\baselineskip}{1em}
Total CCSDT energy in hartree.
}
\end{table*}

\begin{table*}
\caption{
\label{table3}
Convergence of the CC($P$), CC($P$;$Q$)$_{\rm MP}$, and CC($P$;$Q$)$_{\rm EN}$
energies toward CCSDT, where the $P$ spaces consisted of all singles and
doubles and subsets of triples identified during the $i$-FCIQMC, $i$-CISDTQ-MC, or $i$-CISDT-MC
propagations with $\delta\tau = 0.0001$ a.u. and where the corresponding $Q$ spaces
consisted of the triples not captured by the corresponding QMC simulations, for the
reactant (R) and transition-state (TS) structures defining the automerization of cyclobutadiene,
as described by the cc-pVDZ basis set, optimized in the MR-AQCC calculations
reported in Ref. \onlinecite{MR-AQCC}, and for the corresponding activation barrier.
The $i$-FCIQMC, $i$-CISDTQ-MC, and $i$-CISDT-MC calculations preceding the CC($P$) and
CC($P$;$Q$) steps were initiated by placing 100 walkers on the RHF determinant
and the $n_{a}$ parameter of the initiator algorithm was set at 3.
In all post-RHF calculations, the lowest four core orbitals were
kept frozen and the spherical components of $d$ orbitals were employed throughout.
}
\begin{ruledtabular}
\begin{tabular}{lcdddddddddddd}
  &   & \multicolumn{3}{c}{\textrm{\% of triples}} & \multicolumn{3}{c}{\textrm{CC(\textit{P})\footnotemark[1]}} & \multicolumn{3}{c}{\textrm{CC(\textit{P};3)\textsubscript{MP}\footnotemark[1]}} & \multicolumn{3}{c}{\textrm{CC(\textit{P};3)\textsubscript{EN}\footnotemark[1]}} \\ \cline{3-5} \cline{6-8} \cline{9-11} \cline{12-14}
Species & \multicolumn{1}{c}{MC iterations} & \multicolumn{1}{c}{FCI\footnotemark[2]} & \multicolumn{1}{c}{CIQ\footnotemark[3]} &\multicolumn{1}{c}{CIT\footnotemark[4]} & \multicolumn{1}{c}{FCI\footnotemark[2]} & \multicolumn{1}{c}{CIQ\footnotemark[3]} &\multicolumn{1}{c}{CIT\footnotemark[4]} & \multicolumn{1}{c}{FCI\footnotemark[2]} & \multicolumn{1}{c}{CIQ\footnotemark[3]} &\multicolumn{1}{c}{CIT\footnotemark[4]} & \multicolumn{1}{c}{FCI\footnotemark[2]} & \multicolumn{1}{c}{CIQ\footnotemark[3]} &\multicolumn{1}{c}{CIT\footnotemark[4]}\\
  \hline
  R &     0      & & 0 & & & 26.827\footnotemark[5] & & & 4.764\footnotemark[6] & & & 0.848\footnotemark[7] & \\
& 10000 &  0 &  0 &  0 & 25.758& 25.985& 25.484& 4.437&4.535&4.324& 0.696& 0.763& 0.625\\
& 20000 &  2 &  2 &  1 & 22.532& 22.513& 22.462& 3.684&3.621&3.612& 0.496& 0.418& 0.433\\
& 30000 &  6 &  5 &  5 & 17.369& 17.857& 18.880& 2.599&2.676&2.889& 0.230& 0.228& 0.279\\
& 40000 & 16 & 15 & 12 & 11.845& 12.034& 13.834& 1.635&1.649&2.007& 0.092& 0.080& 0.164\\
& 50000 & 31 & 30 & 24 &  6.895&  7.176&  9.202& 0.877&0.913&1.235& 0.022& 0.023& 0.057\\
& 60000 & 52 & 51 & 41 &  3.273&  3.524&  5.205& 0.386&0.417&0.645& 0.001& 0.000& 0.010\\
& 70000 & 72 & 70 & 59 &  1.321&  1.498&  2.594& 0.146&0.170&0.302&-0.003&-0.002&-0.003\\
& 80000 & 85 & 84 & 75 &  0.512&  0.563&  1.181& 0.056&0.060&0.131&-0.001&-0.001&-0.002\\
& \multicolumn{1}{c}{$\infty$} & \multicolumn{3}{c}{100} & \multicolumn{3}{d}{-154.244157\footnotemark[8]} & \multicolumn{3}{c}{---} & \multicolumn{3}{c}{---} \\
  &&&&&\\
  TS & 0 &  & 0 & & & 47.979\footnotemark[5] & & & 20.080\footnotemark[6] & & & 14.636\footnotemark[7] & \\
& 10000&  0 &  0 &  0 & 45.875&46.427&45.777&18.899&19.135&18.037&13.680&13.842&12.665\\
& 20000&  1 &  2 &  1 & 39.577&37.689&39.655&14.220&12.522&13.774& 9.452& 7.793& 8.863\\
& 30000&  5 &  5 &  5 & 30.836&28.405&33.111& 9.660& 7.404&10.798& 5.785& 3.648& 6.651\\
& 40000& 15 & 13 & 13 & 18.976&19.811&23.797& 4.046& 4.313& 6.457& 1.556& 1.661& 3.367\\
& 50000& 31 & 27 & 26 &  9.795& 9.727&12.495& 1.634& 1.488& 2.238& 0.309& 0.243& 0.602\\
& 60000& 52 & 47 & 42 &  3.936& 4.136& 6.217& 0.501& 0.525& 0.886& 0.026& 0.025& 0.105\\
& 70000& 70 & 67 & 60 &  1.491& 1.488& 2.841& 0.173& 0.168& 0.363& 0.003& 0.001& 0.018\\
& 80000& 84 & 82 & 74 &  0.525& 0.591& 1.260& 0.058& 0.065& 0.148& 0.000& 0.000& 0.001\\
& \multicolumn{1}{c}{$\infty$} & \multicolumn{3}{c}{100} & \multicolumn{3}{d}{-154.232002\footnotemark[8]} & \multicolumn{3}{c}{---} & \multicolumn{3}{c}{---} \\
  &&&&&\\
  Barrier & 0 & \multicolumn{3}{c}{0/0} & & 13.274\footnotemark[5] & & & 9.611\footnotemark[6] & & &8.653\footnotemark[7] &\\
& 10000 &  \multicolumn{1}{c}{0/0}  & \multicolumn{1}{c}{0/0}   & \multicolumn{1}{c}{0/0}  &12.624&12.828&12.734&9.075& 9.162&8.605&8.148&8.208&7.555\\
& 20000 &  \multicolumn{1}{c}{2/1}  & \multicolumn{1}{c}{2/2}   & \multicolumn{1}{c}{1/1}  &10.696& 9.523&10.789&6.612& 5.586&6.377&5.620&4.628&5.290\\
& 30000 &  \multicolumn{1}{c}{6/5}  & \multicolumn{1}{c}{5/5}   & \multicolumn{1}{c}{5/5}  & 8.450& 6.619& 8.931&4.431& 2.967&4.963&3.487&2.146&3.999\\
& 40000 &  \multicolumn{1}{c}{16/15}& \multicolumn{1}{c}{15/13} & \multicolumn{1}{c}{12/13}& 4.475& 4.881& 6.252&1.513& 1.672&2.793&0.919&0.992&2.011\\
& 50000 &  \multicolumn{1}{c}{31/31}& \multicolumn{1}{c}{30/27} & \multicolumn{1}{c}{24/26}& 1.820& 1.601& 2.067&0.475& 0.361&0.629&0.181&0.138&0.343\\
& 60000 &  \multicolumn{1}{c}{52/52}& \multicolumn{1}{c}{51/47} & \multicolumn{1}{c}{41/42}& 0.416& 0.384& 0.635&0.073& 0.068&0.151&0.016&0.016&0.060\\
& 70000 &  \multicolumn{1}{c}{72/70}& \multicolumn{1}{c}{70/67} & \multicolumn{1}{c}{59/60}& 0.107&-0.006& 0.155&0.017&-0.001&0.038&0.003&0.002&0.013\\
& 80000 &  \multicolumn{1}{c}{85/84}& \multicolumn{1}{c}{84/82} & \multicolumn{1}{c}{75/74}& 0.008& 0.018& 0.050&0.001& 0.003&0.011&0.001&0.001&0.002\\
& \multicolumn{1}{c}{$\infty$} & \multicolumn{3}{c}{100/100} & \multicolumn{3}{d}{7.627\footnotemark[9]} & \multicolumn{3}{c}{---} & \multicolumn{3}{c}{---}
\end{tabular}
\end{ruledtabular}
\footnotetext[1]{
\setlength{\baselineskip}{1em}
Unless otherwise stated, all energies are reported as errors relative to CCSDT, in millihartree
for the reactant and transition state and in kcal/mol for the activation barrier.
}
\footnotetext[2]{
\setlength{\baselineskip}{1em}
FCI stands for \textit{i}-FCIQMC.
}
\footnotetext[3]{
\setlength{\baselineskip}{1em}
CIQ stands for \textit{i}-CISDTQ-MC.
}
\footnotetext[4]{
\setlength{\baselineskip}{1em}
CIT stands for \textit{i}-CISDT-MC.
}
\footnotetext[5]{
\setlength{\baselineskip}{1em}
Equivalent to CCSD.
}
\footnotetext[6]{
\setlength{\baselineskip}{1em}
Equivalent to the CCSD energy corrected for the effects of $T_{3}$ clusters using
the CCSD(2)$_{\rm T}$ approach of Ref. \onlinecite{ccsdpt2}, which is equivalent to the approximate
form of the completely renormalized CR-CC(2,3) approach of Refs. \onlinecite{crccl_jcp,crccl_cpl},
abbreviated sometimes as CR-CC(2,3),A or CR-CC(2,3)$_{\rm A}$.
\cite{jspp-jctc2012,nbjspp-molphys2017,crccl_jpc,ptcp2007,crccl_ijqc2}
}
\footnotetext[7]{
\setlength{\baselineskip}{1em}
Equivalent to the CCSD energy corrected for the effects of $T_{3}$ clusters using the
most complete variant of the completely renormalized CR-CC(2,3) approach of Refs. \onlinecite{crccl_jcp,crccl_cpl},
abbreviated sometimes as CR-CC(2,3),D or CR-CC(2,3)$_{\rm D}$.
\cite{jspp-jctc2012,nbjspp-molphys2017,crccl_jpc,ptcp2007,crccl_ijqc2}
}
\footnotetext[8]{
\setlength{\baselineskip}{1em}
Total CCSDT energy in hartree.
}
\footnotetext[9]{
\setlength{\baselineskip}{1em}
The CCSDT activation barrier in kcal/mol.
}
\end{table*}

\begin{table*}[h!]
\caption{\label{table4}
Convergence of the CC($P$), CC($P$;$Q$)$_{\rm MP}$, and CC($P$;$Q$)$_{\rm EN}$
energies toward CCSDTQ, where the $P$ spaces consisted of all singles and
doubles and subsets of triples and quadruples identified during the $i$-FCIQMC or $i$-CISDTQ-MC
propagations with $\delta\tau = 0.0001$ a.u. and where the corresponding $Q$ spaces
consisted of the triples not captured by the corresponding QMC simulations,
for the equilibrium and four displaced geometries of the ${\rm H}_{2}{\rm O}$ molecule,
as described by the cc-pVDZ basis set, taken from Ref. \onlinecite{olsen-h2o}.
The $i$-FCIQMC and $i$-CISDTQ-MC calculations preceding the CC($P$) and
CC($P$;$Q$) steps were initiated by placing 100 walkers on the RHF determinant
and the $n_{a}$ parameter of the initiator algorithm was set at 3.
All electrons were correlated and the spherical components of $d$ orbitals were employed throughout.
}
\begin{ruledtabular}
\begin{tabular}{lcdddddddd}
  &  & \multicolumn{2}{c}{\textrm{\% of triples/quadruples}} & \multicolumn{2}{c}{\textrm{CC($P$)\footnotemark[2]}} & \multicolumn{2}{c}{\textrm{CC($P$;$Q$)$_{\rm MP}$\footnotemark[2]}} & \multicolumn{2}{c}{\textrm{CC($P$;$Q$)$_{\rm EN}$\footnotemark[2]}} \\ \cline{3-4} \cline{5-6} \cline{7-8} \cline{9-10}
$R_{\rm O\mbox{-}H}/R_{e}$\footnotemark[1] & MC iterations & \multicolumn{1}{c}{FCI\footnotemark[3]} & \multicolumn{1}{c}{CIQ\footnotemark[4]} & \multicolumn{1}{c}{FCI\footnotemark[3]} & \multicolumn{1}{c}{CIQ\footnotemark[4]} & \multicolumn{1}{c}{FCI\footnotemark[3]} & \multicolumn{1}{c}{CIQ\footnotemark[4]} & \multicolumn{1}{c}{FCI\footnotemark[3]} & \multicolumn{1}{c}{CIQ\footnotemark[4]} \\
  \hline
  1.0 & 0 & \multicolumn{2}{c}{0/0} & \multicolumn{2}{d}{3.725\footnotemark[5]} & \multicolumn{2}{d}{0.887\footnotemark[6]} & \multicolumn{2}{d}{0.325\footnotemark[7]} \\
&10000	& 2/0   & 2/0  &3.291&3.291&0.718&0.718&0.220&0.220\\
&20000	& 4/1   &4/1   &2.874&2.874&0.633&0.629&0.205&0.185\\
&30000	& 6/1	  &5/1	 &2.637&2.637&0.544&0.600&0.143&0.184\\
&40000	& 11/2	&9/2	 &2.052&2.052&0.441&0.471&0.142&0.129\\
&50000	& 13/2	&14/3	 &1.910&1.910&0.390&0.358&0.105&0.095\\
&60000	& 17/3	&18/4	 &1.481&1.481&0.304&0.323&0.087&0.106\\
&70000	& 22/5	&22/5	 &1.238&1.238&0.245&0.249&0.065&0.076\\
&80000	& 27/6	&27/6	 &0.956&0.956&0.207&0.216&0.073&0.082\\
&100000	& 36/10	&35/10 &0.586&0.586&0.127&0.143&0.048&0.065\\
    & $\infty$ & \multicolumn{2}{c}{100} & \multicolumn{2}{d}{-76.241841\footnotemark[8]} & \multicolumn{2}{c}{---} & \multicolumn{2}{c}{---} \\
  &&&&&&&&\\
  1.5 & 0 & \multicolumn{2}{c}{0/0} & \multicolumn{2}{d}{9.922\footnotemark[5]} & \multicolumn{2}{d}{2.704\footnotemark[6]} & \multicolumn{2}{d}{1.021\footnotemark[7]} \\
&10000	&3/1    &3/1   &6.612&6.545&1.393&1.501&0.290&0.434\\
&20000	&8/1    &7/1   &4.068&4.168&0.898&0.799&0.236&0.138\\
&30000	&11/2	  &11/2	 &3.000&3.032&0.613&0.698&0.144&0.248\\
&40000	&16/3	  &17/3	 &1.878&2.207&0.481&0.503&0.231&0.189\\
&50000	&22/4	  &22/4	 &1.465&1.507&0.377&0.366&0.185&0.166\\
&60000	&26/6	  &27/6	 &0.993&0.959&0.254&0.270&0.133&0.152\\
&70000	&31/8	  &33/9	 &0.786&0.706&0.229&0.206&0.133&0.122\\
&80000	&36/10	&38/11 &0.552&0.548&0.186&0.156&0.130&0.091\\
&100000	&46/17	&48/18 &0.259&0.263&0.086&0.086&0.061&0.060\\
    & $\infty$ & \multicolumn{2}{c}{100} & \multicolumn{2}{d}{-76.072227\footnotemark[8]} & \multicolumn{2}{c}{---} & \multicolumn{2}{c}{---} \\
  &&&&&&&&\\
  2.0 & 0 & \multicolumn{2}{c}{0/0} & \multicolumn{2}{d}{22.002\footnotemark[5]} & \multicolumn{2}{d}{3.775\footnotemark[6]} & \multicolumn{2}{d}{-0.581\footnotemark[7]} \\
&10000	&2/0    &2/0   &11.766  &11.803  &1.966&2.189&-0.044&0.200\\
&20000	&7/1    &6/1   &4.172   &4.937   &1.129&1.295& 0.567&0.626\\
&30000	&10/2	  &9/1	 &3.132   &3.788   &0.708&0.683& 0.323&0.160\\
&40000	&14/3	  &13/2	 &1.728   &1.966   &0.603&0.668& 0.436&0.483\\
&50000	&19/4	  &19/4	 &1.123   &1.120   &0.421&0.509& 0.324&0.437\\
&60000	&25/6	  &24/6	 &0.794   &0.719   &0.305&0.221& 0.246&0.156\\
&70000	&30/8	  &30/8	 &0.429   &0.427   &0.129&0.144& 0.094&0.110\\
&80000	&36/11	&35/11 &0.327   &0.293   &0.106&0.103& 0.079&0.082\\
&100000	&47/18	&47/18 &0.107   &0.102   &0.036&0.026& 0.029&0.021\\
    & $\infty$ & \multicolumn{2}{c}{100} & \multicolumn{2}{d}{-75.951635\footnotemark[8]} & \multicolumn{2}{c}{---} & \multicolumn{2}{c}{---} \\
  &&&&&&&&\\
  2.5 & 0 & \multicolumn{2}{c}{0/0} & \multicolumn{2}{d}{22.668\footnotemark[5]} & \multicolumn{2}{d}{-13.469\footnotemark[6]} & \multicolumn{2}{d}{-20.739\footnotemark[7]} \\
&10000	&3/0    &3/0   &18.305  &-3.327  &-1.136&-18.549&-4.962&-21.357\\
&20000	&6/1    &6/1   &5.254   &7.207   & 0.010&  0.448&-0.821& -0.588\\
&30000	&10/2	  &9/2	 &2.278   &2.109   & 0.513&  0.988& 0.298&  0.872\\
&40000	&15/3	  &13/3	 &1.021   &1.170   & 0.304&  0.542& 0.220&  0.490\\
&50000	&22/5	  &17/4	 &0.459   &0.585   & 0.264&  0.287& 0.254&  0.264\\
&60000	&27/8	  &23/6	 &0.340   &0.424   & 0.105&  0.222& 0.096&  0.212\\
&70000	&34/12	&29/9	 &0.133   &0.411   & 0.059&  0.020& 0.054& -0.033\\
&80000	&42/16	&36/13 &0.088   &0.155   & 0.014&  0.052& 0.011&  0.045\\
&100000	&55/28	&49/22 &0.020   &0.027   & 0.013&  0.020& 0.012&  0.020\\
    & $\infty$ & \multicolumn{2}{c}{100} & \multicolumn{2}{d}{-75.920352\footnotemark[8]} & \multicolumn{2}{c}{---} & \multicolumn{2}{c}{---} \\
  &&&&&&&&\\
  3.0 & 0 & \multicolumn{2}{c}{0/0} & \multicolumn{2}{d}{15.582\footnotemark[5]} & \multicolumn{2}{d}{-28.302\footnotemark[6]} & \multicolumn{2}{d}{-35.823\footnotemark[7]} \\
&10000	&3/1    &3/1    &10.165  &12.515  &-2.390&-1.199&-3.945&-2.697\\
&20000	&5/1    &5/1    &4.282   &2.721   &-0.084&-0.690&-0.403&-0.875\\
&30000	&9/2	  &8/2    &1.616   &3.019   & 0.544& 0.357& 0.414& 0.007\\
&40000	&13/3	  &11/3	  &0.969   &0.830   & 0.267& 0.378& 0.199& 0.334\\
&50000	&18/5	  &17/5	  &0.523   &0.400   & 0.251& 0.196& 0.231& 0.184\\
&60000	&24/8	  &22/7	  &0.185   &0.237   & 0.097& 0.093& 0.090& 0.087\\
&70000	&30/12	&28/10	&0.082   &0.128   & 0.039& 0.076& 0.036& 0.075\\
&80000	&36/16	&34/14	&0.030   &0.050   & 0.022& 0.030& 0.021& 0.029\\
&100000	&51/28	&48/24	&0.005   &0.012   & 0.005& 0.008& 0.005& 0.008\\
    & $\infty$ & \multicolumn{2}{c}{100} & \multicolumn{2}{d}{-75.916679\footnotemark[8]} & \multicolumn{2}{c}{---} & \multicolumn{2}{c}{---} \\
\end{tabular}
\end{ruledtabular}
\footnotetext[1]{
\setlength{\baselineskip}{1em}
The equilibrium geometry, $R_{\rm O\mbox{-}H} = R_{e}$, and the geometries that
represent a simultaneous stretching of both O--H bonds by factors of 1.5, 2.0, 2.5, and 3.0
without changing the $\angle$(H--O--H) angle
were taken from Ref. \onlinecite{olsen-h2o}.
}
\footnotetext[2]{
\setlength{\baselineskip}{1em}
Unless otherwise stated, all energies are reported as errors relative to CCSDTQ in millihartree.
}
\footnotetext[3]{
\setlength{\baselineskip}{1em}
FCI stands for \textit{i}-FCIQMC.
}
\footnotetext[4]{
\setlength{\baselineskip}{1em}
CIQ stands for \textit{i}-CISDTQ-MC.
}

\footnotetext[5]{
\setlength{\baselineskip}{1em}
Equivalent to CCSD.
}
\footnotetext[6]{
\setlength{\baselineskip}{1em}
Equivalent to the CCSD energy corrected for the effects of $T_{3}$ clusters using
the CCSD(2)$_{\rm T}$ approach of Ref. \onlinecite{ccsdpt2}, which is equivalent to the approximate
form of the completely renormalized CR-CC(2,3) approach of Refs. \onlinecite{crccl_jcp,crccl_cpl},
abbreviated sometimes as CR-CC(2,3),A or CR-CC(2,3)$_{\rm A}$.
\cite{jspp-jctc2012,nbjspp-molphys2017,crccl_jpc,ptcp2007,crccl_ijqc2}
}
\footnotetext[7]{
\setlength{\baselineskip}{1em}
Equivalent to the CCSD energy corrected for the effects of $T_{3}$ clusters using the
most complete variant of the completely renormalized CR-CC(2,3) approach of Refs. \onlinecite{crccl_jcp,crccl_cpl},
abbreviated sometimes as CR-CC(2,3),D or CR-CC(2,3)$_{\rm D}$.
\cite{jspp-jctc2012,nbjspp-molphys2017,crccl_jpc,ptcp2007,crccl_ijqc2}
}
\footnotetext[8]{
\setlength{\baselineskip}{1em}
Total CCSDTQ energy in hartree.
}
\end{table*}

\clearpage


\end{document}